\documentclass[11pt,letterpaper]{article}

\usepackage[margin=1in]{geometry} 
\usepackage{amsmath,amsthm,amssymb}
\usepackage{graphicx}
\usepackage{tensor}
\usepackage{cancel}
\usepackage{amsmath}
\usepackage{multicol}
\usepackage{graphicx}  
\usepackage{caption}   
\usepackage{soul}
\usepackage{slashed}
\usepackage{subcaption}
\usepackage{subcaption}
\usepackage{hyperref}
\usepackage{physics}

\begin{document}

\begin{titlepage}

\begin{flushright}
arXiv:2510.01106
\end{flushright}

\vskip 2.5cm

\begin{center}
{\Large \bf First Order Axial Perturbation of the Reissner-Nordstr\"{o}m\\
Metric in a Possible Parity-Violating Gravity Background}
\end{center}

\vspace{1ex}

\begin{center}
{\large Abhishek Rout$^{1}$ and Brett Altschul$^{2}$}

\vspace{5mm}
{\sl Department of Physics and Astronomy} \\
{\sl University of South Carolina} \\
{\sl Columbia, SC 29208}\\
\vspace{1.5 mm}
\footnotesize{\tt $^{1}$arout@email.sc.edu\\
$^{2}$altschul@mailbox.sc.edu}
\end{center} 

\vspace{2.5ex}

\medskip

\centerline {\bf Abstract}

\bigskip

We study axial perturbations of Reissner-Nordstr\"{o}m black holes within the general framework of
parity-violating modified gravity theories. We derive the governing equations
for a class of frame-dragging perturbations, focusing on the symmetry structure and radial dependence of the
perturbed metric component, describing its behavior across three distinct regions: near the
singularity ($r \rightarrow 0$), between the inner and outer Reissner-Nordstr\"{o}m horizons ($r_-< r< r_+$),
and in the asymptotic exterior regime ($r \rightarrow \infty$). Using a combination of analytical and
numerical methods,
we analyze the solutions for varying black hole charge-to-mass ratios ($Q/M$) and angular momentum parameters
($l$). Key findings include the suppression of perturbations by the electromagnetic field for higher $Q/M$;
the emergence of radial resonance-like behavior for specific $l$ values; and a high degree of symmetry
for solutions in the extremal limit ($Q/M \sim 1$), attributed to the AdS$_2 \times S^2$ near-horizon geometry.
The WKB approximation is employed to study the high-$l$ regime, revealing quantized radial resonance
modes and singular behavior in the extremal limit. Additionally, we explore the role of boundary conditions
and the possibility of a Chern-Simons field $\Theta$ as the source of the parity violation, showing
that consistency and the behavior of the perturbations under time reversal demand a constant field
(and thus no actually observable Chern-Simons effects) at leading order.
These results provide a basis for further analysis of the stability and dynamical properties of charged
black holes in parity-violating theories, with potential experimental signatures in gravitational wave
observations.

\bigskip

\end{titlepage}

\newpage

\section{Introduction}
\label{sec_intro}

The study of perturbations around highly-symmetric idealized black hole (BH) geometries has been a
cornerstone of gravitational physics, offering critical insights into stability, quasinormal modes (QNMs),
gravitational waves, and potential observational signatures. In the context of charged BHs, the
Reissner-Nordstr\"{o}m (RN) metric can serve as the fundamental unperturbed solution, and
perturbations around it have been extensively explored in General Relativity (GR). Early work by
Regge, Wheeler~\cite{ref-regge}, Zerilli~\cite{ref-zerilli}, and Moncrief~\cite{ref-moncrief} laid
the groundwork for understanding even- and odd-parity perturbations of Schwarzschild and RN BHs,
while later studies extended these analyses to include couplings to additional
fields and higher-order effects~\cite{ref-franciolini,ref-pound2}.

Symmetry is a key topic in fundamental physics, and there is a long history of interest in the possibility
that the apparent symmetries of GR may not hold exactly. Breaking of parity symmetry is an especially
interesting possibility in this regard---and the one that we shall be primarily concerned with here.
In particle physics, parity appears to be conserved in
most low-energy processes; however, the parity violation that arises out of the weak interaction has turned
out to be crucially important to our understanding of the standard model. Although there is no
current experimental
evidence for it, it is also possible that there might be parity violation in the gravitational sector.

In recent decades, modified gravity theories have emerged as a fertile ground for exploring deviations
from GR, motivated by theoretical challenges such as quantum gravity and observational anomalies like
dark energy~\cite{ref-capozziello,ref-clifton,ref-joyce,ref-shankaranarayanan}.
A number of possible forms that parity violation might take in modified gravity theories have been
considered. In Einstein-Cartan theories of gravity, in which spacetime is endowed with a torsion
tensor in addition to its metric, the effective spin sources of the torsion fields are not subject to
any constraint that they should be the same for right- and left-handed source particles~\cite{ref-kibble1}.
From the perspective of test particles,
modified theories that obey the Galileo weak equivalence principle (meaning the
universality of free-falling trajectories) but not the stronger Einstein equivalence principle may
exhibit anomalous parity-violating torques between spinning bodies~\cite{ref-ni1,ref-ni2}.
Some specific models of---and searches for---parity violation in gravitational physics have introduced
the parity violation in conjunction with a broader breaking of local Lorentz
symmetry~\cite{ref-bailey2,ref-battat,ref-muller5,ref-panjwani,ref-hees,ref-kost31,ref-flowers,
ref-kost32,ref-shao3,ref-dong1} in effective field theory. In these Lorentz-violating
field theory models, the
vacuum is endowed with background vectors or tensors with free spacetime indices---which could arise,
for example, as expectation values of vector- or tensor-valued dynamical fields. When the number of
temporal indices on such a background is odd, the resulting dynamics break parity symmetry
In strictly phenomenalistic models (not necessarily
described by local field theories) in which right- and
left-circularly polarized gravitational waves couple to matter differently, there would be parity-violating
imprints left on the cosmic microwave background~\cite{ref-lue,ref-contaldi}. General parameterization
of gravitational parity violation may also feature both velocity and amplitude
birefringence for propagating gravitational waves~\cite{ref-jenks,ref-rout2}, or even
birefringence in the cosmological redshift~\cite{ref-altschul43}. Very recently, another work has
examined QNMs of a Kerr-type solution in parity-violating quadratic gravity~\cite{ref-tahara}.

Among specific modified gravity theories,
Chern-Simons (CS) gravity---a parity-violating extension of GR---has garnered
significant attention, due to its unique coupling between a scalar field $\Theta$ and the Pontryagin
density, which introduces new degrees of freedom and dynamics.
In typical modified gravity theories, additional terms in the action for
the metric field---such as the $\Theta$ term in CS gravity---introduce new dynamics to
BH perturbations.
Previous studies by Jackiw,
Pi~\cite{ref-jackiw5}, Yunes, Sopuerta~\cite{ref-yunes3},
and Alexander~\cite{ref-alexander2}, have examined this theory, including axial perturbations in CS
gravity for Schwarzschild BHs, demonstrating how the scalar field $\Theta$ influences gravitational
waves and stability. In general, this type of
theory might be directly constrained with measurements of gravitomagnetic effects or gravitational
waves~\cite{ref-alexander1,ref-smith1,ref-yunes1,ref-canizares1,ref-yagi1,ref-nakamura1}. When it
was first introduced, it was expected that the gravitational CS term would violate local Lorentz symmetry
as well as parity, since it may be parameterized using a constant background vector contracted with
a combination of Christoffel symbols. In quantum field theory, there is close connection between
discrete symmetries and Lorentz symmetry; it is not possible, in a quantum field theory with a well-defined
$S$-matrix, to violate the combined parity-time reversal-charge conjugation (CPT) symmetry without
also breaking Lorentz invariance~\cite{ref-greenberg}. However, this is no longer the case
in geometric theories of gravity. Although a nontrivial gravitational CS term in the action may not be
written in a Lorentz-invariant form, the apparent Lorentz violation is actually a gauge artifact,
and the theory describes locally-Lorentz-invariant but CPT-violating dynamics~\cite{ref-guarrera}.
This gives the
gravitational CS term a fundamentally different structure from its analogues in particle physics,
including key differences in the structures of the radiative corrections in the two types of
theories~\cite{ref-altschul37}.

This work was initially motivated, in part, by an interest in CS-modified gravity, in its role
as a full dynamical theory of modified gravity. Parity-violating
charged BHs, including RN metrics in CS-modified gravitation, have remained largely
unexplored, leaving a gap in our understanding of how electromagnetic fields and parity violation
could jointly shape perturbation dynamics. Previous studies had examined axial perturbations in CS
gravity for neutral BHs, but the case of charged BHs, particularly in the presence of
nontrivial scalar field profiles, were far less explored. This work looks at BH perturbations in
potentially parity-violating modified gravity theories. However, while we keep a possible
connection the CS gravity in mind, we will proceed with a broader approach, choosing a particular
form of parity-violating perturbation to the metric and analyzing its consequences---focusing,
when possible, on the interplays
between the electromagnetic field, spacetime curvature, and a potential CS coupling.

We begin with a perturbed RN metric featuring an axial perturbation $h_{t\phi}$, which preserves much of
the symmetry of the background while introducing a time-space $dt\,d\phi$ cross term.
The form of the perturbation function $h_{t\phi}=H(r,\theta)$ is decomposed into radial and angular parts,
leading to a
second-order ordinary differential equation (ODE) for the radial component $R(r)$.
We solve this equation analytically in asymptotic regimes (near the singularity, horizons, and 
spatial infinity) and numerically in the bounded region between the inner and outer horizons.
Our setup includes a detailed study of the dependence of solutions on the charge-to-mass ratio $Q/M$ and the
angular momentum parameter $l$, revealing distinct behaviors such as amplitude suppression, resonance
effects, and symmetrization in the extremal limit.

Consideration of just one type of perturbation is a significant restriction, but a general computational
analysis of all possible perturbations would be extremely complicated, and so we leave
much of it for further work. The particular metric perturbation form we have selected, time independent
and with axial symmetry, preserves one temporal
and one angular Killing vector from the unperturbed RN metric, and so it retains the maximal degree of possible
symmetry consistent with nontrivial angular and parity-violating effects.
We initially considered more general perturbations, but without a significant degree of residual symmetry,
the results were difficult to interpret.
Obviously, considering
time-dependent perturbations---representing oscillatory and radiating QNMs---is also of great interest; it
is something we plan to take up in the future, but it lies beyond our current scope. In fact, for our numerical
computations, we further limit our primary attention to one specific region of the spacetime---that which
lies between the two horizons of the RN metric.  There is quite a rich structure to the perturbations
just within this coordinate region, as we shall demonstrate, although exploration of other portions of
the spacetime, with different characteristic boundaries and boundary conditions, is also a topic for
future work.

It will turn out that, for reasons ultimately related to time reversal symmetry, that the $h_{t\phi}$
perturbation considered here is not really amenable to nontrivial CS dynamics.
We shall come to this conclusion based on a general symmetry argument. However, for completeness we may
also calculate explicitly the form that the CS scalar field $\Theta$ would need to take to support
our perturbative framework,
finding that consistency demands it to be constant unless additional symmetry-breaking terms are introduced.
For high-$l$ regimes, we will employ the Wentzel–Kramers–Brillouin (WKB) approximation to derive radially
quantized mode profiles  and discuss their physical implications. Our results not only extend the
understanding of RN BH perturbations in modified gravity but also highlight the rich interplay
between charge, angular momentum, and boundary conditions in shaping the dynamics of such systems.

This paper is structured as follows. Section~\ref{sec2} outlines the theoretical framework and
perturbation equations, ending with a discussion of the remaining spacetime symmetries of the
perturbed solutions. Then section~\ref{sec3} presents analyses in certain special limits. Section~\ref{sec4}
introduces the parameters involved in our numerical calculations of the field profiles for perturbations
around the RN metric. The results of these calculations are presented in the next sections---with the
dependence of the field profiles on the charge-to-mass ratio and boundary conditions in section~\ref{sec5},
and the dependences of the amplitude of the pertubations in section~\ref{sec-ampl}. We then return to
discussion of some analytical issues. The CS scalar $\Theta$ is explicitly shown to vanish at leading order in
the perturbations (as expected) in section~\ref{sec-Theta}, confirming why our calculations
do not actually depend on $\Theta$ or its derivatives. In section~\ref{sec-WKB},
we introduce the WKB approximation
to help understand certain resonance behavior that was pointed out in section~\ref{sec-ampl}. And
section~\ref{sec-AdS} discusses the special symmetries of the extremal RN spacetime and their relevance to our
calculations. In section~\ref{sec:observations}, we discuss potential experimental implications of
our results, particularly how they may be useful in helping to distinguish between different modified gravity
models. Section~\ref{sec-concl} concludes with a discussion of further implications and future directions.
More analytical details of the calculation of perturbed tensor forms and of the WKB approximation are
given in appendices.

\section{Axial Perturbations}
\label{sec2}

To investigate the gravitational response of a charged BH in a potentially parity-violating
modified gravity theory, we
consider perturbations around the standard static RN background, which is an exact
solution of the Einstein field equations with or without a CS term in the action.
We begin with the assumption that the metric perturbation takes the form
\begin{equation}
    g_{\mu\nu} = \bar{g}_{\mu\nu} + \epsilon h_{\mu\nu},
    \label{eq-metric-perturbation}
\end{equation}
where $\bar{g}_{\mu\nu}$ is the background RN metric~\cite{ref-reissner}, $\epsilon$
is a small dimensionless parameter characterizing the perturbation strength, and $h_{\mu\nu}$ represents
the metric perturbation. The perturbation satisfies the constraint
$|\epsilon h_{\mu\nu}| \ll |\bar{g}_{\mu\nu}|$, ensuring we are remaining within the linear perturbation
regime where higher-order terms in $\epsilon$ can be neglected.

\subsection{Physical Motivation for $t$-$\phi$ Perturbations}

We specifically consider perturbations where only the $h_{t\phi}$ component is non-zero. This choice is
physically motivated by several considerations:
\begin{itemize}
    \item \textbf{Axial Symmetry Preservation:} The RN background possesses full spherical
    symmetry. Introducing perturbations that break this symmetry while preserving axial symmetry (rotational
    invariance about a preferred axis) represents a natural first step. The $t$-$\phi$ component represents
    the minimal coupling between the temporal and azimuthal directions that maintains axial symmetry.
    \item \textbf{Frame-Dragging Effects:} The $h_{t\phi}$ perturbation describes a cross term
    $dt\,d\phi$ in the line element, which physically corresponds to \textit{frame-dragging}---the phenomenon
    where a rotating mass (or in this case, a perturbation that mimics rotational effects) drags the
    surrounding spacetime along with it along a revolving congruence of trajectories.
    Even in the absence of actual rotation of the source, parity violation may generate
    effective frame-dragging through metric perturbations.
    \item \textbf{Sensitivity to Parity Violation:} 
    The $t$-$\phi$ perturbation corresponds to an odd-parity (axial) perturbation that couples directly to
    the parity-violating sector of the theory, making it a good probe for detecting novel
    symmetry-breaking physics.
    \item \textbf{Minimality:} Among all possible off-diagonal metric perturbations, $h_{t\phi}$ represents
    the minimal deviation from spherical symmetry that is capable of introducing new physical effects. Other
    off-diagonal components would either break axial symmetry ($h_{t\theta}$, $h_{\phi\theta}$) or
    represent different physical phenomena ($h_{tr}$, $h_{r\phi}$).
\end{itemize}

\subsection{Mathematical Form of the Perturbed Metric}

With only $h_{t\phi}$ non-zero, the perturbed metric takes the explicit form
\begin{equation}
    ds^2 = - f(r)dt^2 
           + \frac{1}{f(r)}dr^2 
           + r^2(d\Omega^2)
           + 2\epsilon h_{t\phi}(r,\theta)\,dt\,d\phi,
    \label{eq-perturbed-metric}
\end{equation}
where $f(r)$ is called the ``lapse function'' and is (in natural units with $c = G = 1$)
\begin{equation}
    f(r) = 1 - \frac{2M}{r} + \frac{Q^2}{r^2};
\end{equation}
here $M$ is the mass, and $Q$ the total charge, of the BH. The perturbation function
$h_{t\phi}(r,\theta)$ may depend on both radial and angular coordinates, reflecting its axial symmetry.
The time independence ($\partial_t h_{t\phi} = 0$) of the perturbation is physically motivated; since the
background RN metric is static (time-independent and non-rotating), we consider
stationary perturbations that preserve this static character while introducing frame-dragging effects only
through the off-diagonal $t$-$\phi$ term. This approach allows us to study how parity-violating
corrections might modify the
properties of static charged BHs without introducing any complicating time-dependent dynamics.

\subsection{Physical Interpretation}

The $dt\,d\phi$ term in the line element has several potentially important physical implications.
\begin{itemize}
    \item \textbf{Angular Momentum Exchange:} The perturbation $h_{t\phi}$ can be interpreted as generating
    an effective angular momentum distribution in spacetime, even if the background solution carries no
    physical angular momentum. In the context of CS gravity, something like this might arise from parity-violating
    interactions between the gravitational field and an external scalar field $\Theta$.
    \item \textbf{Geodetic Precession Modification:} Test particles orbiting the BH will experience
    modified geodetic precession due to the $dt\,d\phi$ term. This provides potential observational
    signatures in precision measurements of satellite orbits or pulsar timing.
   \item \textbf{Perturbation of Horizon Structure:} While the background horizon at
   $r = r_+ = M + \sqrt{M^2 - Q^2}$ remains unchanged to first order, the $h_{t\phi}$ perturbation
   modifies the geometry in its vicinity, potentially affecting properties like the surface gravity
   and horizon area at higher orders.
\end{itemize}

Because a major motivation for this work was CS gravitation, we shall discuss the structure of
that specific theory.
The Einstein-Hilbert action for CS-modified gravity represents a compelling extension of GR
that captures potential parity-violating gravitational effects. We consider primarily consider a non-dynamical
formulation of the CS theory, where the key scalar field $\Theta$ is prescribed externally rather than being
determined dynamically from the action. 
This $\Theta$ is particularly relevant in scenarios where parity
violation emerges from fundamental physics beyond GR, such as in string theory
compactifications or loop quantum gravity. The action for the theory is
\begin{equation}
    \mathcal{S} = \frac{1}{16\pi G}\int d^4x \,\sqrt{-g} \left[\mathcal{R} 
    + \frac{\Theta}{4}(^{\star}\mathcal{R}\mathcal{R}) 
    - \frac{1}{4}F_{\mu\nu}F^{\mu\nu}
    + \mathcal{L}_{\text{mat}}\right],
    \label{eq-actionS}
\end{equation}
where we have:
\begin{enumerate}
    \item \textbf{Standard Einstein-Hilbert term} ($\mathcal{R}$)\textbf{:} Represents the
    conventional geometrodynamics of spacetime.
    \item \textbf{CS correction} $\left(\frac{\Theta}{4}\,^{\star}\mathcal{RR}\right)$\textbf{:}
    This term introduces parity violation through the Pontryagin density $^{\star}\mathcal{RR}$, where
    $^{\star}R$ is the dual of the Riemann tensor. Physically, this term breaks the symmetry between how
    left-handed and right-handed gravitational waves couple, potentially leaving imprints in the
    polarization patterns of the cosmic microwave background or gravitational wave observations.
    \item \textbf{Maxwell electromagnetic term} ($-\frac{1}{4}F_{\mu\nu}F^{\mu\nu}$)\textbf{:}
    Represents the standard dynamics of electromagnetic fields minimally coupled to gravity.
    \item \textbf{Matter Lagrange density} ($\mathcal{L}_{\text{mat}}$)\textbf{:} Accounts for additional
    matter fields that may be present in the system.
\end{enumerate}

The non-dynamical nature of $\Theta$ means it is treated as an external field with prescribed spacetime
dependence, rather than being determined by an equation of motion. This simplification is particularly useful
for studying observational constraints on parity violation in gravity without introducing additional
complicating dynamical degrees of freedom. (If $\Theta$ were dynamical, it would correspond to a
gravitational-axion-like field.)

Varying the action with respect to the metric tensor $g_{\mu\nu}$ yields the modified field equations,
\begin{equation}
    G_{\mu\nu} + C_{\mu\nu} = 8\pi T_{\mu\nu}.
    \label{eq-fieldeq}
\end{equation}
Here, $G_{\mu\nu} = R_{\mu\nu} - \frac{1}{2}g_{\mu\nu}R$ is the standard Einstein tensor, and 
$C_{\mu\nu}$ is the four-dimensional Cotton tensor, a geometric object that emerges from the CS
modification. Mathematically, it takes the form
\begin{equation}
        C^{\mu\nu} = -\frac{1}{2\sqrt{-g}}\left[\Theta_{,\alpha}\epsilon^{\alpha\beta\gamma(\mu}\nabla_{\gamma}
        R^{\nu)}_{\ \beta} + \Theta_{,\alpha\beta}^{\ \ \ \ \star}R^{\beta(\mu\nu)\alpha}\right],
        \label{eq-cottontensor}
\end{equation}
where parentheses around indices denote symmetrization. Physically, $C_{\mu\nu}$ encodes the parity-violating
corrections to Einstein's equations and satisfies $\nabla_\mu C^{\mu\nu} = 0$ as required by diffeomorphism
invariance.
Naturally, $T_{\mu\nu} = T_{\mu\nu}^{\text{(EM)}} + T_{\mu\nu}^{\text{(mat)}}$ is the total stress-energy
tensor, with the electromagnetic contribution being given by
\begin{equation}
        T_{\mu\nu}^{\text{(EM)}} = \frac{1}{4\pi}\left(F_{\mu\alpha}F_{\nu}^{\ \alpha}
        -\frac{1}{4}g_{\mu\nu}F_{\alpha\beta}F^{\alpha\beta}\right).
        \label{eq-EMtensor}
\end{equation}

The perturbative expansion of these field equations around the RN background spacetime (detailed in
Appendix~\ref{app:perturbations}) reveals how the CS corrections modify both the constraint equations and
the field equations for metric perturbations. Considering the perturbative expansion of the field equations,
we note that the Cotton tensor $C_{\mu\nu}$ vanishes to first order in perturbation theory for the specific
$t$-$\phi$ component we consider. This occurs because the Cotton tensor depends on derivatives of the scalar
field $\Theta$ and the background Riemann tensor; for our stationary perturbation around the static RN
background, these contributions cancel at linear order. Alternatively, we shall shortly see that this
vanishing is also a consequence of the time reveral properties of our $h_{t\phi}$ metric perturbation.
It follows that the RN metric is a solution of the field equations
with linearized perturbations. (In fact, it remains an exact solution of the modified
theory~\cite{ref-alexander2,ref-nashed}.)
Consequently, the Chern-Simons scalar field $\Theta$
does not appear explicitly in the linearized field equation for $h_{t\phi}$, simplifying the analysis while
still capturing modifications to the gravitational response through the perturbed geometry.

With this simplification, the field equation for the $t$-$\phi$ component specifically takes the form
\begin{equation}
    \frac{r^2 f(r)}{2} \partial^2_r H + \frac{2M}{r} H + \frac{1}{2} \partial^2_{\theta} H
    - \cot{\theta}\, \partial_{\theta} H = 0,
    \label{eq-H-PDE}
\end{equation}
where $H(r,\theta) \equiv h_{t\phi}(r,\theta)$. We shall endeavor to solve this partial differential
equation using separation of variables, decomposing $H(r,\theta)$ into radial and angular components to
obtain solutions that describe the frame-dragging perturbation around the charged BH.

Using the separation method [that is, assuming the ansatz $H(r,\theta) = R(r)\vartheta(\theta)$],
we see that the partial differential field equations separate into two ordinary differential 
equations (ODEs),
\begin{equation}
    \begin{aligned}
        \frac{fr^2}{2}\frac{d^2 R(r)}{dr^2} + \left[\frac{2M}{r} - \lambda\right]R(r) & = 0\\
        \frac{d^2\vartheta(\theta)}{d\theta^2} - 2\cot{\theta}\frac{d\vartheta(\theta)}{d\theta}
        + 2\lambda\vartheta(\theta) & = 0.
    \end{aligned}
\end{equation}
The second differential equation, for $\vartheta$, is a standard one, which can be solved by
Legendre polynomials. The regular solutions will be of the form,
\begin{equation}
    \vartheta(\theta) = P_{l}(\cos{\theta}),
\end{equation}
where $\lambda = l(l+1)$, with $l$ a non-negative integer. Thus, the
equation for the radial part becomes
\begin{equation}\label{eq-radial}
    \frac{d^2 R(r)}{dr^2} + \left[\frac{4M}{fr^3} - \frac{2l(l+1)}{fr^2}\right]R(r) = 0,
\end{equation}
which has the form of a Schr\"{o}dinger equation with a highly singular potential. In fact,
with the lapse function $f$, \eqref{eq-radial} is
actually a form of Heun's equation~\cite{ref-ronveaux,ref-slavyanov}, with known exact solutions in terms
of special functions; however, for analytic solutions, we shall only consider approximate forms near the
horizons, leaving the equation to be solved numerically in the intermediate region.

\subsection{Symmetries of the Perturbations}

The symmetries of the perturbations we are looking at are important to understanding their nature
and potential effects. Having laid out the nature of the perturbation solutions, it is now possible
to examine the residual spacetime symmetry structure left behind after the perturbation modifies the
RN metric.

The background RN metric has an obvious $O(3)\times Z_{2}$---rotation, reflection, and
time reversal---symmetry. The inclusion
of a frame-dragging $dt\,d\phi$ term in the line element breaks some of this symmetry. A nonzero
$h_{t\phi}$ term gives rise to a
Lense-Thirring-like precession effect (which is equivalent, as noted above, to a transfer of angular
momentum from the background to a test particle) that is manifestly not invariant under time reversal.
Test particles approaching radially from spatial infinity are entrained into orbits with positive
or negative $\dot{\phi}$ (depending on the sign of $h_{t\phi}$. So for the perturbed metrics
we are considering, the $Z_{2}$ time reversal symmetry (which took $dt\rightarrow -dt$) is lost
as a central factor of the spacetime symmetry group---although a time-reversing transformation may
still be a symmetry if it is accompanied by an additional sign change from the spatial part of the
transformation.

The particular choice of a nonzero $h_{t\phi}$ also breaks part of the rotational symmetry of
the problem. Selecting perturbations only to this component of the metric maintains the rotation
symmetry about the $z$-axis of the BH; the remaining rotations do not preserve $\phi$, but since they
merely change $\phi$ by a constant, they take $d\phi\rightarrow d\phi$.

The trickiest part of the symmetry structure---as befits a study of possibly parity-violating effects---is
with the spatial reflections. We shall consider first the reflection planes containing the $z$-axis
(for the moment setting aside reflections across the $xy$-plane).
A nonzero $h_{t\phi}$ breaks the reflection symmetries
across the vertical planes, since such reflections take $d\phi\rightarrow -d\phi$. However, this negative
sign would be compensated for if there were also a time reversal, giving a net
$dt\,d\phi\rightarrow dt\,d\phi$. The result is a group that is isomorphic to $O(2)$---a semidirect
product $SO(2)\rtimes Z_{2}$---but which acts differently on spacetime, compared with the usual $O(2)$
group of origin-fixing rigid motions of the $xy$-plane. Specifically, any spatial reflection must be
accompanied by a time inversion; in other words, the improper transformations must also be
nonorthochronous.

The inclusion of possible reflections across the $xy$-plane complicates things further. These reflections
introduce the possibility of another $Z_{2}$ factor. However, the way they combine with the symmetry
transformations discussed above depends on $l$. If $l$ is even,
so that $H(r,\theta)=H(r,\pi-\theta)$, a reflection across the $xy$-plane does not introduce another sign,
and the full symmetry group is a further direct product of the $SO(2)\rtimes Z_{2}$ from above with another
copy of $Z_{2}$. In contrast, if $l$ is odd, $H(r,\theta)=-H(r,\pi-\theta)$, and reflection across the
horizontal plane produces another minus sign. The result is another semidirect product,
$[SO(2)\rtimes Z_{2}]\rtimes Z_{2}$.

In both cases, the total symmetry group is actually
isomorphic to $O(2)\times Z_{2}$, but the group of symmetries
embeds differently inside this abstract group depending on whether $l$ is even or odd. The key feature is
that the discrete contributors to a
symmetry transformation must produce an overall even number of negative signs. When $l$ is even,
the negative signs are associated with time reversal and any reflection across a vertical plane. When $l$
is odd, the minus signs come from these two classes of discrete transformations, and also from reflections
through across the horizontal $xy$-plane.

There is something else to be learned from from this discussion of the spacetime symmetries. Although
this analysis was initiated as a exploration of parity-violating gravity involving a CS term, a
frame-dragging $h_{t\phi}$ perturbation of the metric is not actually expected to be associated with
a nonzero CS term. The CS term---in the action or the field equations---is odd under the combined
action of parity and time reversal. When the non-dynamical CS term, with $\Theta$ prescribed as a linear
function of the time coordinate, the term is odd only under parity, while even under time reversal (and
even under charge conjugation). However, the $dt\,d\phi$ structure associated with $h_{t\phi}$ is always
odd under time reversal. Thus, we expect (and shall indeed find) that perturbations $h_{t\phi}=H(r,\theta)$
will not generate a nonzero contribution to the field equations.

\section{Limiting Behaviors}
\label{sec3}

\subsection{Asymptotic Behavior as $r \rightarrow \infty$}
\label{sec-asymptotic}

In the limit $r \rightarrow \infty$, far from the BH horizon, the gravitational influence of the
central mass becomes weak. For the radial equation governing $R(r)$, the term $4M/r^3$ decreases as
$\mathcal{O}(r^{-3})$, while the angular momentum barrier term $-2l(l+1)/r^2$ decreases more slowly as
$\mathcal{O}(r^{-2})$. Therefore, at sufficiently large distances, the dominant behavior is governed by
the approximate equation,
\begin{equation}
    \frac{d^2 R(r)}{dr^2} - \frac{2l(l+1)}{r^2}R(r) = 0.
    \label{eq-R-large-r}
\end{equation}
This is an Euler-Cauchy type differential equation, which admits power-law solutions of the form
$R(r) = r^{\alpha}$. Substituting this ansatz yields the characteristic equation
\begin{equation}
    \alpha(\alpha-1) - 2l(l+1) = 0,
\end{equation}
with solutions
\begin{equation}
    \alpha_{\pm} = \frac{1 \pm \sqrt{1 + 8l(l+1)}}{2}.
\end{equation}

This kind of solution describes the behavior of perturbations in the weak-field region, far from the BH.
This regime is physically relevant for a couple reasons:
\begin{itemize}
    \item \textbf{Observational Relevance:} Astronomical observations of BHs typically probe regions well
    outside the horizon, where $r \gg M$. The asymptotic behavior informs us about observable
    frame-dragging effects at large distances.
    \item \textbf{Matching to External Solutions:} While the BH interior requires careful treatment of the
    horizon region, the exterior solution must connect smoothly to the weak-field regime. The power-law
    decrease ensures the perturbation remains small at infinity, consistent with isolated systems in
    asymptotically flat spacetime.
\end{itemize}

The general asymptotic solution is a linear combination,
\begin{equation}
    R(r) \approx A_{\infty}r^{\alpha_{+}} + B_{\infty}r^{\alpha_{-}}.
\end{equation}
For $l \geq 0$, $\alpha_{+} > 0$ and $\alpha_{-} < 0$. The growing solution ($\propto r^{\alpha_{+}}$) is
non-physical as it diverges at spatial infinity, violating the requirement that perturbations should vanish
far from the source. Therefore, we impose the boundary condition $A_{\infty} = 0$ to select the physically
admissible spatially decaying solution,
\begin{equation}
    R(r) \approx \frac{B_{\infty}}{r^{|\alpha_{-}|}}, \quad \text{as } r \rightarrow \infty,
    \label{eq-asymptotic-decay}
\end{equation}
with the negative $\alpha_{-}$.

While the mathematical limit $r \rightarrow \infty$ represents a mathematically idealized spatial infinity,
in physical applications this corresponds to distances much larger than the characteristic scale of the system
($r \gg M$). This asymptotic analysis
provides the correct functional form of the perturbation in the weak-field region surrounding the BH.
Already in the RN metric, the spacetime outside the outer horizon is not a vacuum, because of the pervasive
electric field. So it is reasonable to have modifications to the spacetime that also persist into the far
field of the perturbed solutions.
However, the decreasing power-law behavior ensures that the perturbation remains small compared to the
background metric at large distances, and the solution can be smoothly matched to an external environment, like
the interstellar medium or a cosmological background, if one is present. For numerical implementations,
the asymptotic form provides guidance for boundary conditions at finite but large radii, avoiding
nonphysical divergences.
In practice, one would implement boundary conditions at some finite but large radius $r_{\text{max}} \gg M$
based on the solved asymptotic behavior, ensuring the solution remains regular throughout the computational
domain.

\subsection{Behavior Near the Singularity at $r \rightarrow 0$}

As we approach the central singularity at $r = 0$, the RN lapse function $f(r) = 1 - 2M/r + Q^2/r^2$ is
dominated by the charge term:
\begin{equation}
    f(r) \approx \frac{Q^2}{r^2}, \quad \text{as } r \rightarrow 0.
    \label{eq-f-near-singularity}
\end{equation}
This behavior reflects the physical assumption that the electric charge $Q$ is effectively concentrated near
the central singularity, as expected for a point charge in classical electrodynamics. In the context of
GR, the RN solution represents the exact exterior metric of a spherically symmetric charged
mass, with the charge and mass parameters $M$ and $Q$ completely characterizing the central object. While
a complete physical description of the singularity requires quantum gravity, our classical analysis near
$r = 0$ can provide insights into how perturbations behave in the strong-field region close to the central
object.
In general, the approximation $f(r) \approx Q^2/r^2$ as $r \rightarrow 0$ is valid in the regime where the
coordinate $r$ is much smaller than both the mass and charge scales ($r \ll M, |Q|$). In this region,
The electric repulsion encoded in $Q^2/r^2$ dominates over the gravitational attraction $-2M/r$, indicating
that near the singularity, electromagnetic effects become increasingly important relative to purely
gravitational ones.

Substituting the approximation \eqref{eq-f-near-singularity} into the radial ODE yields
\begin{equation}
   \frac{d^2 R(r)}{dr^2} + \left[\frac{4M}{Q^2 r} - \frac{2l(l+1)}{Q^2}\right]R(r) = 0.
   \label{eq-R-small-r}
\end{equation}
As $r \rightarrow 0$, the term $4M/(Q^2 r)$ dominates the coefficient of $R(r)$, indicating that near
the singularity, the mass parameter $M$ continues to influence the perturbation dynamics despite the
dominance of electromagnetic terms in the lapse function.

Using a power-law ansatz $R(r) = r^{\alpha}$ in the limit $r \rightarrow 0$, the lowest power
term $r^{\alpha-2}$ dominates, giving the indicial equation,
\begin{equation}
    \alpha(\alpha - 1) = 0,
\end{equation}
with solutions $\alpha = 0$ and $\alpha = 1$. Thus, to leading order, the general solution near the
singularity behaves as
\begin{equation}
    R(r) \approx A_0 + B_0 r,
    \label{eq-leading-singularity}
\end{equation}
where $A_0$ and $B_0$ are integration constants. The constant term ($\alpha = 0$) dominates as
$r \rightarrow 0$, suggesting that physically regular perturbations should approach a finite, non-zero value
at the singularity rather than diverging.

Interestingly, equation \eqref{eq-R-small-r} without the $l$-dependent term can be transformed into a version of
Bessel's equation. The exact solutions are linear combinations of $\sqrt{r}J_1(\sqrt{4Mr}/Q)$ and
$\sqrt{r}N_1(\sqrt{4Mr}/Q)$, where $J_1$ and $N_1$ are Bessel functions of the first and second kind.
Both solutions are regular at $r = 0$ due to the $\sqrt{r}$ prefactor. The power-law forms $A_0$ and $B_0 r$
correspond to the leading terms in the series expansions of these Bessel function solutions.

For a physically admissible perturbation, we require that $R(r)$ remains finite as $r \rightarrow 0$.
This selects the constant solution $R(r) \approx A_0$ as the physically relevant leading behavior.
This regularity condition is analogous to the requirement on finite curvature invariants in physically
reasonable space-times, although we note that the background metric itself is singular at $r = 0$.

In summary, the analysis near $r = 0$ reveals that perturbations remain finite at the central singularity,
with the constant solution $A_0$ representing the dominant behavior. This result is consistent with treating
$Q$ as the total charge concentrated at the origin, as in the classical point charge model embedded within
GR.

\subsection{Behavior Near the Horizons}
\label{sec-near-horizons}

The RN spacetime has two horizons, at two radii
\begin{equation}
r_{\pm}=\frac{1}{2}\left(2M\pm\sqrt{4M^{2}-4Q^{2}}\right).
\end{equation}
The outer horizon in the event horizon of the spacetime, while the inner horizon is a Cauchy horizon.
Between the two surfaces, the variable $r$ is timelike.
At these horizons, the function $f(r)$ becomes zero and hence the ODE again becomes singular. So
we shall also analyze the behavior near the horizons. As $r \rightarrow r_{\pm}$ we can write
approximately $f(r)$ as
\begin{equation}
    f(r) = f'(r_{\pm})(r - r_{\pm}),
\end{equation}
where $f'(r)$ is the derivative of $f(r)$ with respect to $r$. Using this in the ODE, we get
\begin{equation}
    \frac{d^2 R(r)}{dr^2} + \left[\frac{4M}{(f')(r - r_{\pm})r_{\pm}^3} -
    \frac{2l(l+1)}{(f')(r - r_{\pm})r_{\pm}^2}\right]R(r) = 0;
    \label{eq-R-horizon}
\end{equation}
the derivatives are, of course, evaluated at $r_{\pm}$.
The ODE~\eqref{eq-R-horizon} can be written in the more unified form,
\begin{equation}
    \frac{d^2 R(r)}{dr^2} + \frac{\beta_{\pm}}{(r - r_{\pm})} = 0,
\end{equation}
where
\begin{equation}
    \beta_{\pm} = \frac{4M}{f'(r_{\pm})r_{\pm}^3} - \frac{2l(l+1)}{f'(r_{\pm})r_{\pm}^2}.
\end{equation}
This is a second-order linear ODE with a simple pole at $r = r_{\pm}$. To solve it, we can use a
Frobenius series expansion. We assume a solution of the form
\begin{equation}
    R(r) = (r - r_{\pm})^s\sum^{\infty}_{n=0}\,a_n(r - r_{\pm})^n.
\end{equation}
For simplicity, if we focus on the leading order behavior near $r = r_{\pm}$. On substitution of the
ansatz into the ODE we get
\begin{equation}
    s(s - 1)(r - r_{\pm})^{s-2} + \beta_{\pm}(r - r_{\pm})^{s-1} = 0.
\end{equation}
The dominant term near $r = r_{\pm}$ is $(r - r_{\pm})^{s-2}$. Factoring that out and using the fact that
$(r - r_{\pm})^{s-2} \neq 0$ as we approach $r \rightarrow r_{\pm}$, we get a relation
\begin{equation}
    s(s-1) + \beta_{\pm}(r  -r_{\pm}) = 0
\end{equation}
Now, as $r \rightarrow r_{\pm}$ the term, $\beta_{\pm}(r - r_{\pm})$ becomes negligible and hence, to
leading order, the characteristic equation is once again of the form $s(s - 1) = 0$, implying
\begin{equation}
     s = 0\hspace{0.5cm}\text{or}\hspace{0.5cm} s = 1.
\end{equation}
Hence the general solution near $r = r_{\pm}$ is supposed to look like.
\begin{equation}
    R(r) \sim A_{\pm} + B_{\pm}(r - r_{\pm})
\end{equation}
Once again, these are approximate forms, when the actual solutions involve Bessel functions (in this case,
Bessel functions of a shifted radial argument).
Again, for the solution to remain finite and nonzero near $r = r_{\pm}$, we require the constant
$B_{\pm} = 0$; in that case, the dominant behavior near a horizon, $r \rightarrow r_{\pm}$, is
\begin{equation}
    R(r) \approx A_{\pm}.
\end{equation}

While the linear terms $B_{\pm}(r - r_{\pm})$ formally vanishes at the corresponding horizons, their derivative
$\partial_r R \approx B_{\pm}$ obviously need not. 
The numerical solutions in the following sections demonstrate that $R(r)$ frequently exhibits nonzero
slopes near $r_{\pm}$,  implying $B_{\pm} \neq 0$ in what seem to be the most physically relevant scenarios. 
This behavior reflects the perturbative energy flux across the horizons, which is generically non-vanishing
for dynamical perturbations, involving mixed time-space components of the metric. 
The boundary condition $|R(r \to r_{\pm})| < \infty$ remains satisfied regardless, since
$(r - r_{\pm}) \to 0$ at the boundaries, rendering both $A_{\pm}$ (constant) and $B_{\pm}$ (linear) terms
admissible. 
Any apparent tension between our subsequent numerical calculations and the Frobenius analysis
may be largely resolved by noting that the numerical solutions will be evaluated at $r = r_{\pm} \pm \epsilon$,
slightly away from the horizons, where the $B_{\pm}$ contributions can be definitely nonzero.

While our current analysis focuses on stationary perturbations ($h_{t\phi}$ independent of time),
the horizon analysis should be useful for obtaining an understanding of the more general case of dynamical
perturbations in the future. In particular, the horizons play a fundamental role in determining the
QNMs of BHs---complex-frequency oscillations that characterize the ringdown phase
following gravitational disturbances. These modes satisfy specific radiation boundary conditions: purely
incoming waves at the innermost horizon and purely outgoing waves at spatial
infinity~\cite{ref-chandrasekhar1,ref-steinhauer}. Although our
stationary analysis does not directly compute QNMs, the horizon behavior we derive can form the foundation
for such eventual calculations. The regularity conditions at $r = r_{\pm}$ will ensure that perturbations
remain finite at the horizons, a prerequisite for well-defined QNM boundary conditions in time-dependent analyses.

\subsection{Stability Considerations and the Extremal Limit}

The charge-to-mass ratio $Q/M$ plays a crucial role in determining the stability and physical properties of
RN BHs. Classical GR imposes the bound $|Q| \leq M$ for BH solutions, with the extremal case
$|Q| = M$ representing a critical threshold at which all the mass energy of the BH arises from its electrical
properties.

The relative sizes of $M$ and $|Q|$ are key to the determination of whether an ordinary RN configuration is
stable. For $|Q| < M$, the BH possesses two distinct horizons at $r_\pm = M \pm \sqrt{M^2 - Q^2}$.
As $|Q|/M \rightarrow 1$, these horizons coalesce ($r_+ \rightarrow r_- \rightarrow M$), resulting in an
extremal BH with vanishing surface gravity ($\kappa = 0$). The inner (Cauchy) horizon at $r_-$ is known to
be classically unstable due to the phenomenon of mass inflation~\cite{ref-poisson}, where linear perturbations
become unbounded in the vicinity of $r_-$, regardless of how small an initial perturbation may be.
This instability has significant implications for the cosmic censorship hypothesis and the predictability
of physics inside BHs.

From a thermodynamic perspective, extremal BHs ($|Q| = M$) have zero Hawking temperature (because of the
vanishing surface gravity $\kappa$) and vanishing entropy in the classical limit, although quantum
corrections modify this picture~\cite{ref-zaslavskii}. The third law of BH thermodynamics suggests that
reaching exact extremality requires an infinite number of steps, making truly extremal BHs physically
unrealizable. Furthermore, quantum effects such as Schwinger pair production near charged BHs can lead to
charge neutralization, driving $|Q|/M$ away from extremality~\cite{ref-gibbons}.

In the context of CS modified gravity, the stability analysis should account for additional
parity-violating contributions. Previous studies~\cite{ref-yunes:2009hc,ref-konno:2009kg} have shown that
CS corrections can modify the effective potential for metric perturbations, potentially altering the
stability boundaries. For charged BHs, the coupling between electromagnetic and gravitational
perturbations introduces additional modes whose stability depends on both $Q/M$ and the CS coupling strength.

These considerations related to the stability of time-dependent perturbations will inform our choices of the
numerical paramters for our computational investigations of the stationary case.
We consider a range of $Q/M$ values spanning from the nearly neutral regime ($Q/M = 0.01$)
to near-extremal cases ($Q/M = 0.95$ or $0.99$). This allows us to explore how frame-dragging perturbations
behave across the stability landscape:
\begin{itemize}
    \item For $Q/M \ll 1$, the solutions approximate the Schwarzschild case with small electromagnetic
    corrections.
    \item As $Q/M$ increases toward unity, the horizons approach each other, potentially amplifying
    perturbation effects in the inter-horizon region.
    \item Near the extremal limit, numerical challenges arise due to the vanishing of $f'(r_\pm)$,
    requiring careful treatment of boundary conditions.
\end{itemize}
The avoidance of exactly extremal values ($Q/M = 1$) in our analysis reflects both the physical
unrealizability of such configurations and the numerical difficulties associated with coincident horizons.
Our results for $Q/M = 0.99$ provide insight into near-extremal behavior while remaining within the
regime where traditional perturbative methods and numerical techniques remain valid.

\subsection{Boundary Conditions for Bound-Like States}

As seen from the previous subsections, the radial profile $R(r)$ of the metric perturbation adheres to certain
trends as we move close to the salient points of the RN metric, at $r = 0$, $r_-$, and $r_+$ and also
asymptotically very far away from the outer horizon, $r \rightarrow \infty$.

Now, in the region between the Cauchy horizon and the event horizon,
the solution is expected to be finite and may
exhibit oscillations, like a quantum-mechanical wave function between two turning points of a potential.
However, to determine the exact solution we need to define the boundary conditions at these inner and outer
horizon boundaries. Exactly at the boundaries $f(r)  = 0$, and the ODE becomes singular. So we
shall, as noted above, define the boundaries for our numerical calculations to be very near the horizons,
rather than lying exactly at them. The boundary condition~\cite{ref-li} that is natural for such a case is
simply
\begin{equation}
   |R(r \to r_\pm)| < \infty
\end{equation}
which simply keeps the the solution finite within the boundaries (or the horizons).
The other boundary condition typical of bound states that needs to be satisfied is
\begin{equation}
    |R(r \rightarrow \infty)| = 0.
\end{equation}
However, for scattering-like states, the boundary condition is determined by the asymptotic behavior and
again merely satisfies
\begin{equation}
    |R(r \rightarrow \infty)| < \infty.
\end{equation}

The constant terms will predominate in computations starting from $r = r_{\pm} \pm \epsilon$, and it is
easy to see that the boundary conditions that $R(r)$ remain finite at the two horizons are satisfied
for any choices of $A_{\pm}$ and $B_{\pm}$; these give the value and the first radial derivative of $R(r)$
in the vicinities of $r_{\pm}$. [The full Bessel function solutions near the horizons are likewise necessarily
finite, since they only contain additional terms with higher powers of $(r-r_{\pm})$, which vanish as the
horizons are approached.

Moreover, we already imposed the boundary condition that the solution not be growing as
$r \rightarrow \infty$ in order to find the solution
\begin{equation}
    R(r) \approx \frac{B_{\infty}}{r^{|\alpha_{-}|}}.
\end{equation}
Since the other solution grew at large distances, there do not seem to be any perturbations (of the type
we have been considering) that have scattering-like behavior in the far field regime.

\section{Numerical Implementation and Boundary Treatment}
\label{sec4}

We shall now move from analytical to numerical analyses of the structures of the perturbation functions.
We will be using the Wolfram Mathematica numerical ODE solver \texttt{NDSolve} to solve the ODE in
three regions, distinctly defined as $0 < r < r_-$, $r_- < r < r_+$, and $r_+ < r < \infty$. For now,
we shall focus on the bound-like states to be obtained in the region between the horizons. Subsequently, we
will analyze the behavior in the external and internal spaces.
  
The ODE~\eqref{eq-radial} for the radial function $R(r)$ becomes singular at the horizons $r = r_\pm$,
where $f(r) = 0$, making the equation numerically stiff and potentially unstable near these points. To
obtain physically reliable numerical solutions, we implement a small offset from the horizons by
defining the numerical integration domain as $r \in [r_- + \epsilon, r_+ - \epsilon]$, where $\epsilon$ is a
small positive parameter. For our computations, we use $\epsilon = 10^{-6}$ in our natural units. This choice
ensures we remain sufficiently far from the singular horizons while maintaining proximity to the physical
domain of interest between the inner and outer horizons.

\subsection{Boundary Conditions and $\epsilon$-Dependence}

At each of the boundaries of the integration region, we impose a Dirichlet boundary condition,
\begin{eqnarray}
     R(r_- + \epsilon) & = & A_-\\
     R(r_+ - \epsilon) & = & A_+,
\end{eqnarray}
where $A_\pm$ are finite real constants. One of the values may be rescaled to 1, due to the linearity of the
equation.
The boundary conditions at $r_\pm \pm \epsilon$ describe the behavior of the frame-dragging perturbation
in the vicinities of the horizons. We require that the perturbation does not diverge as it approaches a
horizon, which would be physically unacceptable for a regular perturbation. 

\begin{figure}[h]
    \centering
    \includegraphics[width=0.9\linewidth]{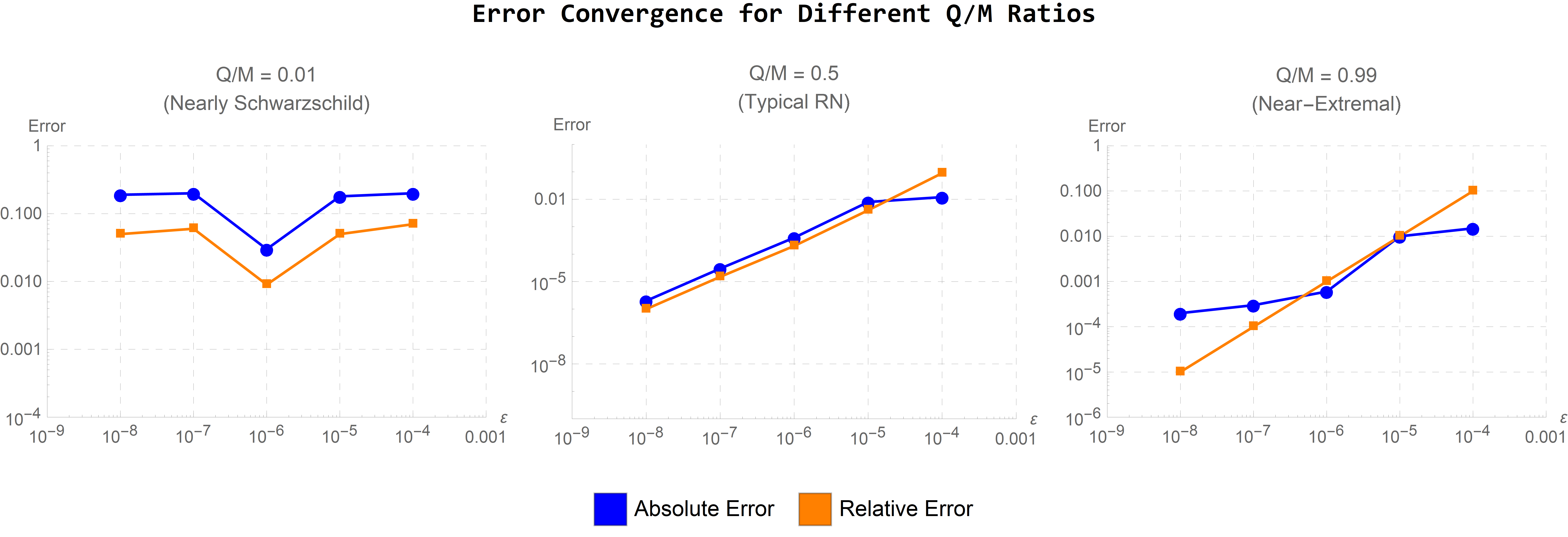}
    \caption{Convergence tests showing absolute and relative errors versus boundary offset
    $\epsilon$ for three representative charge-to-mass ratios: nearly neutral ($Q/M = 0.01$),
    intermediate ($Q/M = 0.5$), and near-extremal ($Q/M = 0.99$). The optimal balance between accuracy
    and computational cost occurs at $\epsilon = 10^{-6}$ across all regimes.}
    \label{fig:error_conv_comp}
\end{figure}

The value $\epsilon = 10^{-6}$ was selected based on systematic numerical convergence tests across the
full parameter space of interest. To determine the optimal distance from the horizons for imposing
boundary conditions, we computed solutions for decreasing values of $\epsilon = 10^{-4}$, $10^{-5}$, $10^{-6}$,
$10^{-7}$, and $10^{-8}$ for various charge-to-mass ratios $Q/M$ (ranging from nearly neutral
$Q/M = 0.01$ to near-extremal $Q/M = 0.99$) and angular parameters $l$. Comparisons were made with
the results for $\epsilon=10^{-9}$, a value which was small enough that the solutions had essentially
completely stabilized and so could be taken to be closely representative of the exact (albeit
computationally intensive) solution. The results for three representative 
instances of $Q/M$ are shown in figure~\ref{fig:error_conv_comp}; the other parameters used in these
specific plots were $l=4$ and $\gamma$---the boundary values parameter defined in \eqref{eq-gamma-def}
below---of 1.

The detailed analysis
revealed that the error behavior is significantly influenced by the underlying BH geometry, with three
distinct regimes emerging:

\textbf{Nearly neutral regime ($Q/M \approx 0.01$):} In this limit approaching the Schwarzschild case,
the inner horizon lies very close to the central singularity at $r = 0$. The error exhibits a pronounced
U-shaped profile with a clear minimum at $\epsilon = 10^{-6}$. For $\epsilon > 10^{-5}$, the boundaries are
placed too far from the horizons, failing to capture the near-horizon physics accurately. For
$\epsilon < 10^{-7}$, numerical instabilities arise from integrating too close to the $r \to 0$ singularity,
causing errors to increase again. The optimal $\epsilon = 10^{-6}$ strikes a delicate balance between
these competing effects.

\textbf{Moderate charge regime ($Q/M = 0.5$):} For typical RN BHs with
well-separated horizons, the error decreases monotonically as $\epsilon$ becomes smaller. Here, the inner
horizon is sufficiently far from the singularity, and the equations remain well-behaved arbitrarily close
to both horizons. In this regime, smaller $\epsilon$ consistently yields better accuracy, with relative
errors reaching $\sim 10^{-6}$ at $\epsilon = 10^{-8}$. This ideal convergence behavior confirms that our
numerical method is fundamentally sound.

\textbf{Near-extremal regime ($Q/M = 0.99$):} As the BH approaches extremality, the two horizons
coalesce and $f'(r_{\pm})$ becomes very small, rendering the differential equations extremely stiff.
The error decreases rapidly from $\epsilon = 10^{-4}$ to $10^{-6}$, but further reduction to
$\epsilon = 10^{-7}$ and $10^{-8}$ yields diminishing returns; error improvement slows dramatically from
a factor of $\sim 10$ per decade to only $\sim 1.5$ per decade. This plateau reflects the
limitations of double-precision numerics when applied to the highly singular near-horizon geometry,
which in the extremal limit exhibits an emergent $\mathrm{AdS}_2 \times S^2$ structure with enhanced
conformal symmetry.

Across all three regimes, $\epsilon = 10^{-6}$ emerges as a robust and conservative choice. For nearly
neutral BHs, it sits precisely at the minimum of the U-shaped error curve. For moderate charges,
while smaller $\epsilon$ would give marginally better accuracy, $\epsilon = 10^{-6}$ already provides
relative errors below $0.1\%$. For the numerically challenging near-extremal case, it represents the point
beyond which further refinement yields negligible improvement while risking increased stiffness and numerical
instability. The consistency of solutions across this wide range of physical parameters, with relative errors
at $\epsilon = 10^{-6}$ remaining below $1\%$ in all cases, confirms that our results capture the true
behavior of the system rather than numerical artifacts arising from the boundary treatment.

\subsection{Boundary Condition Parameter $\gamma$}

To characterize the relationship between boundary conditions at the two horizons, we introduce the ratio 
\begin{equation}
\label{eq-gamma-def}
    \gamma = \frac{A_-}{A_+},
\end{equation}
where $A_\pm$ are the amplitudes at $r_\pm \pm \epsilon$, respectively. At the most basic level,
this characterizes how the eventual solutions we shall find connect the the fully
interior and fully exterior solutions in the regions $r<r_{-}$ and $r>r_{+}$. The $r<r_{-}$ solution may have
a complicated variation in that region; however, as discussed in section~\ref{sec-near-horizons},
it approaches the constant $A_{-}$ as $r$ nears $r_{-}$. The factor $\gamma$ describes how this
value relates to the value $A_{+}$ of the exterior solution at the outer, $r=r_{+}$ boundary---and,
therefore, the relative strength of the entire exterior solution, including the long-distance
tail, \eqref{eq-asymptotic-decay} from section~\ref{sec-asymptotic}.

Moreover, it is also possible to think of the parameter $\gamma$ in a somewhat different fashion,
more than simply characterizing the boundary values used
in the numerical integration---sort of an inverse of that description, in fact.
It can provide a useful way of characterizing how the solution in the
intermediate region between the two horizons is governed by its connections to
the fully interior and fully exterior solutions
of the perturbation equation.
Each version of the differential equation governing the perturbation in the
$r_{-}<r<r_{+}$ region admits a family of
solutions. Specifying the ratio of the amplitudes near the inner and outer
horizons picks out one member of this family. In this sense,
$\gamma$ parametrizes different global perturbative solutions that interpolate
between the interior region ($r<r_-$), the inter-horizon region
($r_-<r<r_+$), and the exterior spacetime ($r>r_+$). Whichever of $A_{\pm}$ is larger
pulls the intermediate solution away from zero at its corresponding endpoint.

From this perspective, varying $\gamma$ corresponds to changing how strongly
the intermediate-zone solution is influenced by the boundary data at the inner
horizon relative to the outer horizon. Smaller values of $\gamma$ correspond
to solutions whose characteristics are primarily determined by the exterior behavior and
therefore connect more directly to the asymptotically decaying outer
solutions. Larger values of $\gamma$, on the other hand, emphasize the
contribution from the inner horizon and correspond to solutions that are more
strongly tied to the deep interior structure of the spacetime.

Thus, $\gamma$ serves as a parameter that labels different global solutions of
the perturbation equation, each representing a different way in which the
intermediate solution connects the interior and exterior regions of the
BH spacetime. The introduction of $\gamma$ as a
parameter enhances the flexibility of the numerical setup, enabling a detailed exploration of 
a larger solution space.
So we have the combination of the parameters $\gamma$, $Q / M$, and the previously introduced $l$
to vary to explore the different
kinds of ODE solutions within the bounded region $r_- < r < r_+$.

\section{Solutions in $r_-<r<r_+$}
\label{sec5}

\subsection{Varying $Q / M$ Ratios}

\begin{figure}[h]
    \centering
    \fbox{\includegraphics[width=0.95\textwidth]{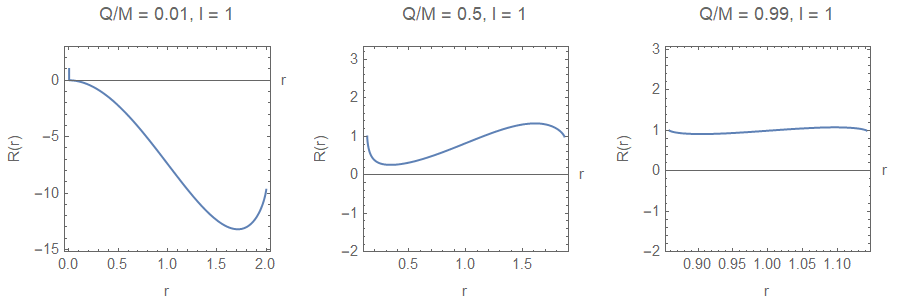}}
    \caption{This plot shows the solutions for three different $Q/M$ ratios when  $l = 1$. On the left is the
    case for a Schwarzschild BH, where $Q \sim 0$; in the center is the case of a typical
    RN BH, for which $Q / M = 0.5$; and on the right is the near-extremal
    RN case, where $ Q / M \sim 1$. Note that in the true extremal limit, $Q/M=1$,
    there is no region between the horizons, because they degenerate to $r_{+}=r_{-}$.}
    \label{fig:plot}
\end{figure}

This subsection contains plots showing the radial solution profiles for those three different $Q/M$ ratios
that describe clearly distinct unperturbed RN BHs. The value of the ratio $\gamma$, characterizing the
difference in $R(r)$ at the two ends of the region is taken to be $\gamma=1$---equality at the two
edges---except in the $l=1$ nearly Schwarszchild case, in which specifying $\gamma$ leads to computational
instability, since one boundary is nearly at the $r=0$ curvature singularity.
The plots show $r$ measured
in units of $M$---so that in the nearly-Schwarszchild case, for example, the plots extend from $r_{-}=0$
to $r_{+}=2M$.
Since the ODE for $R(r)$ is linear, $R(r)$ is plotted in arbitrary units, with the integrations
starting and and ending at inner and outer boundary values of $R(r_{\pm}\pm\epsilon)=1$. The plots cover
several different values of the angular parameter $l$, starting with $l=1$ in Figure~\ref{fig:plot}.

Figure~\ref{fig:plot} shows clear differences in the solutions for the three different $Q/M$ regimes.
The Schwarzschild case ($ Q \sim 0$) exhibits a highly asymmetric behavior in $R(r)$, with a significant
variation in amplitude. As the charge increases, the solution becomes more symmetric relative to the inner
and outer horizons; and in the extremal RN case ($Q/M \sim 1$), the function $R(r)$ approaches
a nearly constant form. This trend suggests that increasing the charge stabilizes the radial function,
reducing spatial variations in
the solution. This could be attributed to the presence of an inner horizon in charged BHs, which shrinks
the physical region of interest and modifies the effective potential experienced by the perturbation.
The transition from the Schwarzschild case to extremality may also reflect the increasing role of the
Coulomb repulsion, which influences the propagation of perturbations. In subsequent plots, we shall analyze
how increasing $l$ affects these trends and whether similar symmetrization occurs for higher angular momenta.

\begin{figure}[h]
    \centering
    \fbox{\includegraphics[width=0.95\textwidth]{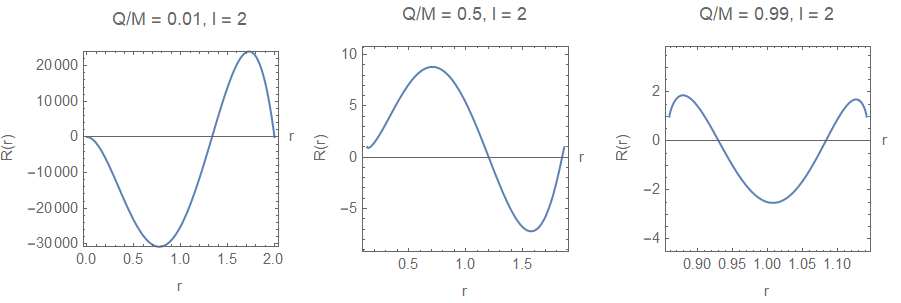}}
    \caption{This plot shows the solution for the three different $Q/M$ ratios when $l = 2$.}
    \label{fig:plot_l2}
\end{figure}

Figure~\ref{fig:plot_l2} illustrates the radial function $R(r)$ for different values of $Q/M$
when the angular parameter
is $l=2$. We can observe a number of key features. The function is increasingly oscillatory in nature.
Compared to the case $l = 1$, the radial function exhibits more oscillations, and this is expected from
experience with quantum mechanics and other problems in BH perturbation theory. Higher $l$ values introduce
more angular nodes in the solution, which must also be accompanied by greater radial variations.

Once again, there are clear effects of the charge on the amplitude of the oscillations. For the
Schwarzschild case $Q \sim 0$, $R(r)$ oscillates with a large amplitude, indicating a significant
variation in the radial function. As $Q/M$ increases to $0.5$, the oscillations persist, but the
amplitude (relative to the boundary values) is massively reduced. In the extremal case $Q/M \sim 1$, $R(r)$
becomes nearly symmetric and has a very small amplitude, similar to what was observed for $l = 1$. This
again suggests that the presence of charge smoothens the radial profile and limits its variation,
and the behavior aligns with the physical intuition that the presence of an inner horizon in the RN
spacetime alters the effective potential, with one result being damping fluctuations in the radial
function. Compared to the $l = 1$ cases, the solutions in all three $Q/M$ regimes are more symmetric
for $l=2$. Evidently, the higher $l$ value introduces more structure in the solution, due to the
larger centrifugal-like term present in the wave equation.

\begin{figure}[h]
    \centering
    \fbox{\includegraphics[width=0.95\textwidth]{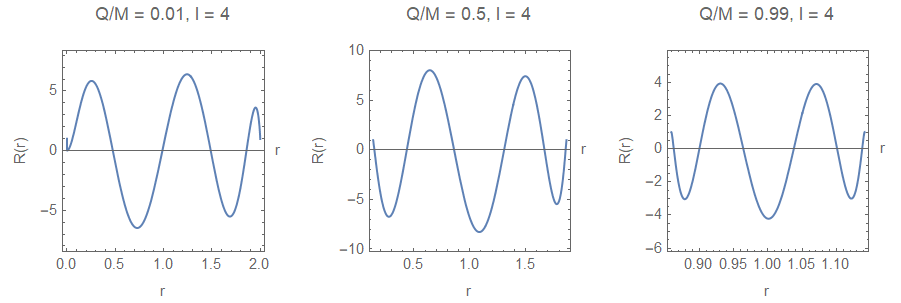}}
    \caption{This plot shows the solution for the three different $Q/M$ ratios when $l = 4$.}
    \label{fig:plot_l4}
\end{figure}

Many of these trends will continue to become more pronounced as $l$ is further increased.
Figure~\ref{fig:plot_l4} shows the radial function $R(r)$ for varying ratios of $Q/M$ when $l = 4$.
Uponn comparison with the plot for $l=2$, we see that the function $R(r)$ again exhibits
increasingly oscillatory behavior, which is typical for the radial components of solutions whose angular
parts are spherical harmonics or similar functions. By $l=4$, the amplitude suppression with increasing
$Q/M$ has largely disappeared; the three profiles are all similar in amplitude. However, the increasing
symmetry is still present, although already by $Q/M=0.5$, the curve is extremely symmetric,
and for the extremal case, it is difficult to see any asymmetry at all.

\begin{figure}[h]
    \centering
    \fbox{\includegraphics[width=0.95\textwidth]{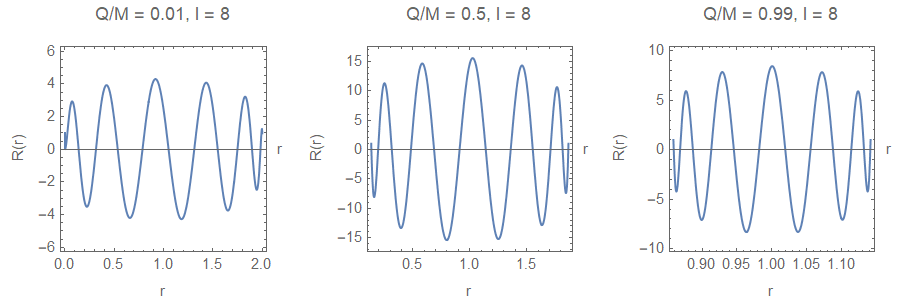}}
    \caption{This plot shows the solution for the three different $Q/M$ ratios when $l = 8$.}
    \label{fig:plot_l8}
\end{figure}

Continuing to increase the angular parameter, Figure~\ref{fig:plot_l8} shows the radial function for $l=8$.
The three curves are starting to look qualitatively very similar, although the extremal case still does show
the greatest degree of symmetry. The previous trend relating the amplitude to $Q/M$ has completely disappeared.
In fact, now the greatest  amplitude for the function is obtained when we are in the intermediate RN regime
with $Q/M=0.5$.

\begin{figure}[h]
    \centering
    \fbox{\includegraphics[width=0.95\textwidth]{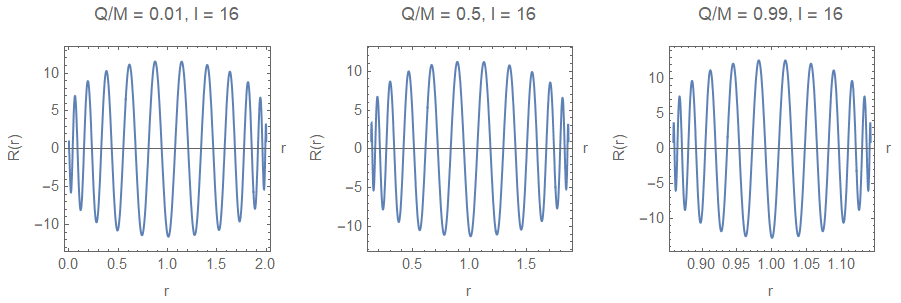}}
    \caption{This plot shows the solution for the three different $Q/M$ ratios when $l = 16$.}
    \label{fig:plot_l16}
\end{figure}

As $l$ has increased, we have seen an expected increase in the number of radial nodes in $R(r)$. Now we
shall  delve into what happens when $l$ is taken to be quite large. In this regime, the angular
quantum number $l$ comes to determine almost completely the complexity and structure of the solutions.
For very high $l$ the solutions to the Schr\"{o}dinger-like ODE become dominated by the centrifugal potential
term, which scales as 
$l(l+1)/r^{2}$. This centrifugal term becomes extremely large for high $l$, effectively overshadowing other
terms in the equation (such as those depending on $Q/M$). As a result, the solutions for different 
$Q/M$ values start to look very similar, because $l$ is becoming the only relevant parameter.
Figure~\ref{fig:plot_l16} shows the solution for $l = 16$.

As expected, we see that for high $l$, the solutions tend to oscillate rapidly in $r$, due to the strong
influence of the centrifugal potential. The oscillations are so rapid and pronounced that the differences
caused by varying $Q/M$ become negligible in comparison; the solutions for $Q/M=0.01$, $Q/M=0.5$, and
$Q/M=0.99$ appear almost identical. In physical systems, high $l$ values correspond to states with very high
angular momentum. In such states, the particle (or wave) is pushed far away from the center due to the
centrifugal force, and the small-$r$ behavior of $f(r)$ (which has a $Q^{2}/r^{2}$ divergence) or the detailed
form of the the central potential in the Schr\"{o}dinger-like equation (which depends on $Q/M$) become far
less significant. The convergence of solutions for different parameter values to a similar pattern when $l$
is large is a common phenomenon in quantum mechanics and wave dynamics, where high angular momentum states
are less sensitive to the details of the central potential near $r=0$.

\subsection{Varying $\gamma$ for Different $Q/M$}

\begin{figure}[h]
\begin{tabular}{ccc}
\includegraphics[width=0.3\textwidth]{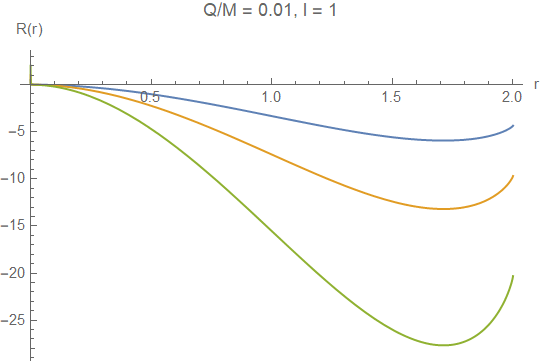} &
\includegraphics[width=0.3\textwidth]{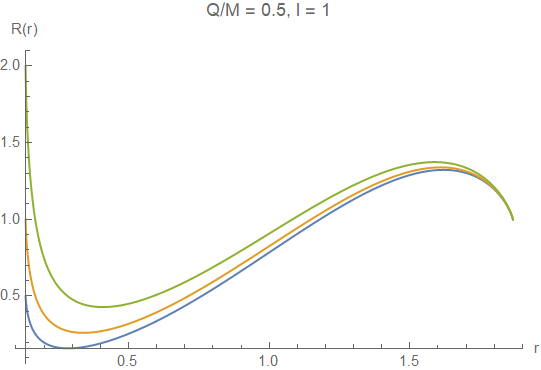} &
\includegraphics[width=0.35\textwidth]{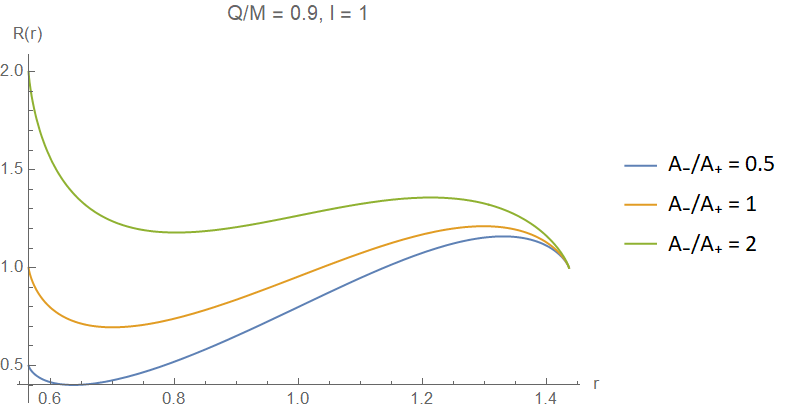}
\end{tabular}
\caption{Plots of the radial function for three different $\gamma$ values in the ratio $1:2:4$. From
left to right are the three cases $Q/M = 0.01$ (nearly Schwarzschild), $Q/M = 0.5$ (typical RN), and 
$Q/M = 0.9$ (nearly extremal RN), with angular parameter $l=1$.
For the center and right graphs, the colors indicate $\gamma=0.5$ (blue),
$\gamma=1$ (green), and $\gamma=2$ (orange). For the left-hand graph (Q/M=0.01), the gamma values are
in the same ratios, but their actual values are much smaller, for reasons of computational
necessity.}
\label{fig:gamma1}
\end{figure}

The $\gamma$ parameter changes the boundary conditions used by \texttt{NDSolve} to obtain the solution for
$R(r)$ in the region between the Cauchy and event horizons. In this subsection, we shall look at the
solutions for the radial function for the three cases when $\gamma < 1$, $\gamma > 1$ and $\gamma = 1$,
all once again as a further function of the angular parameter $l$. The normalization of the plots is
$A_{+}=1$, except for the left graph in Figure~\ref{fig:gamma1}, which was chosen differently for
illustrative purposes.

As we have seen in the previous subsection (Figure~\ref{fig:plot}), the solutions for $l = 1$ vary
considerably depending on $Q/M$. The different values of $\gamma$ create a further spectrum of solutions,
as we adjust the boundary conditions. For the Schwarzschild case ($Q/M \sim 0$), $R(r)$ veers negative
before partially turning around. Once again, specifying a $\mathcal{O}(1)$ value for $\gamma$
in this case leads to
problems, because the inner boundary at $r_{-}$ is so close to $r=0$. However, it is still possible to have
the $\gamma$ values be in the ratio $1:2:4$.

As the $Q/M$ ratio increases, the charge term starts to play a significant role, leading to a different
behavior for $R(r)$ compared to the nearly neutral case. The solution shows a more complex behavior,
indicating the influence of both mass and charge. For $Q/M=0.5$, there is a notable degree of symmetry for
all three values of $\gamma=0.5$, $1$, and $2$, with the solutions lying close together except around
the inner boundary $r_{-}$. As once again we reach the extremal case ($Q/M \sim 1$), the three curves with
these values of $\gamma$ separate again, possibly indicating stronger influence of the electromagnetic charge
as seen in Figure~\ref{fig:gamma1} .

We may now delve into a fuller examination of the effects of changing $\gamma$, which controls
the relative amplitude of the perturbations at the inner and outer horizons.
For $\gamma=0.5$, the boundary condition at the inner horizon is smaller than at the outer horizon.
For intermediate $Q/M$ ratios, the perturbations are primarily influenced by the outer horizon value,
leading to simpler solutions that do no vary much depending on $\gamma$. This is a trend that we
shall continue to see for small and intermediate values of $Q/M$ at higher $l$.
In contast, for high $Q/M$ ratios (e.g., 0.9), the influence of the inner horizon becomes more
pronounced, leading to more complex behavior in $R(r)$.

For $\gamma=1$, with equal boundary at both horizons, the solutions are the ones discussed in the previous
subsection. The solutions for $R(r)$ are balanced between the influences of the inner and outer horizons.
For moderate $Q/M$ ratios (e.g., 0.5), this balance leads to a smoothly symmetric transition in the behavior
of $R(r)$ between the horizons.

Taking $\gamma=2$, the boundary condition at the inner horizon is more significant compared to the
outer horizon. For the largest value of $\gamma$ we used with low $Q/M$ ratios, this led to more pronounced
computational instabilities near the inner horizon; these may or may not be indicative of a physical
tendency toward instability. For high $Q/M$ ratios, the strong influence of the inner horizon can lead to
significant oscillations and complex behavior in $R(r)$.

\begin{figure}[h]
\begin{tabular}{ccc}
\includegraphics[width=0.3\textwidth]{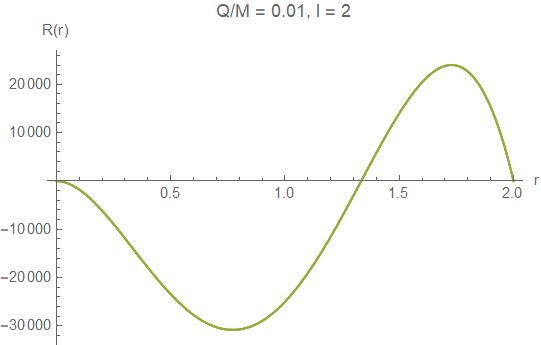} &
\includegraphics[width=0.3\textwidth]{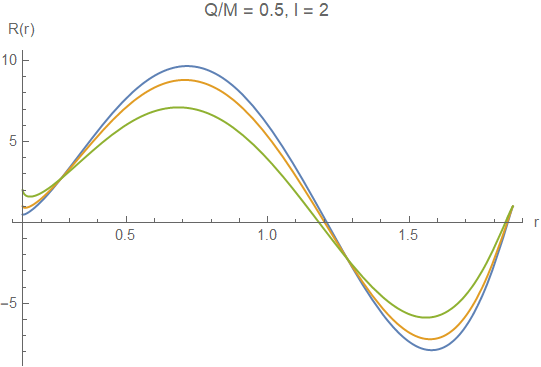} &
\includegraphics[width=0.35\textwidth]{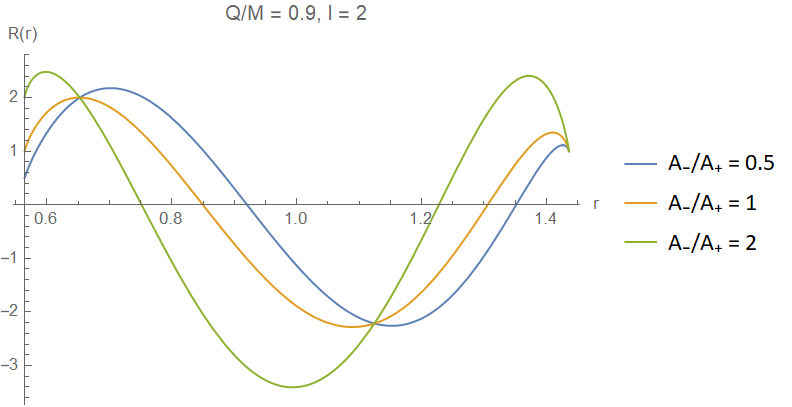}
\end{tabular}
\caption{This plot shows the solutions for the radial functions for the three different $Q/M$ regimes
and $\gamma$ values, all for $l=2$.
All three graphs show $\gamma=0.5$ in blue, $\gamma=1$ in green, and $\gamma=2$ in orange.}
\label{fig-gamma-l2}
\end{figure}

We will now explore the spectra of solutions obtained by varying $\gamma$ for higher $l$ values.
Figure~\ref{fig-gamma-l2} shows the solutions obtained for different $\gamma$ values when we take $l = 2$. 
As expected, the transition from dipolar($l = 1$) perturbations to quadrupolar perturbations ($l = 2$) leads
to more complex, high-wave-number spatial oscillations.

Here, we can observe that the solutions for different $\gamma$ values appear almost identical when we have
$Q/M \sim 0$. Because the overall amplitudes are quite large, modest differences in the inner and outer
boundary values hve very little qualitative effect on the shapes of the curves.
Alternatively, this can perhaps be attributed to the fact that when we have nearly a neutral BH, the
inner horizon is very close to the singularity, and its influence on the perturbations is minimal.
The outer horizon (or in the case of an exact Schwarszchild spacetime, the only horizon) dominates the
behavior of the perturbations. The boundary conditions at the inner horizon have little effect because the
inner horizon is not significantly influencing the perturbation dynamics. 

When $Q/M$ grows greater, the effect of both the horizons become more pronounced, and the curves for
different $\gamma$ values are clearly separated, although they still are quite similar in overall shape.
Turning to the case when we have extremality ($Q/M \sim 1$), the solutions are oscillatory---each oscillating
the same number of times---but they differ quite a bit in their other attributes. This is due to the fact
that for $Q/M = 0.9$,  the inner horizon is very close to the outer horizon, and the perturbations are
strongly sensitive to the boundary conditions at both horizons.

Changing $\gamma$ alters the relative influence of the inner and outer horizons, leading to different
oscillation patterns in $R(r)$. For $Q/M=0.9$, the effective potential is highly sensitive to the boundary
conditions, leading to different oscillation frequencies and amplitudes for different $\gamma$ values.
When $\gamma = 2$, the inner horizon’s dominance leads to stronger oscillations, while for
$\gamma = 0.5$, the outer horizon’s dominance results in somewhat milder oscillations. The oscillations
in $R(r)$ probably correspond to energy being exchanged between the gravitational and electromagnetic fields.
The differences in oscillation patterns for different $\gamma$ values suggest that the energy
dissipation mechanisms depend strongly on the boundary conditions.
 
\begin{figure}[h]
\begin{tabular}{ccc}
\includegraphics[width=0.3\textwidth]{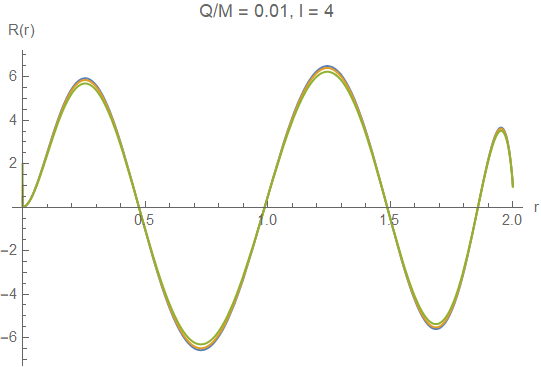} &
\includegraphics[width=0.3\textwidth]{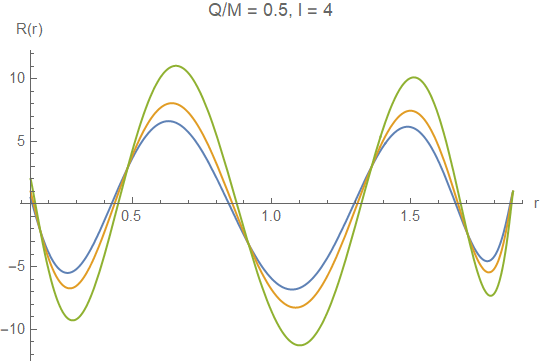} &
\includegraphics[width=0.35\textwidth]{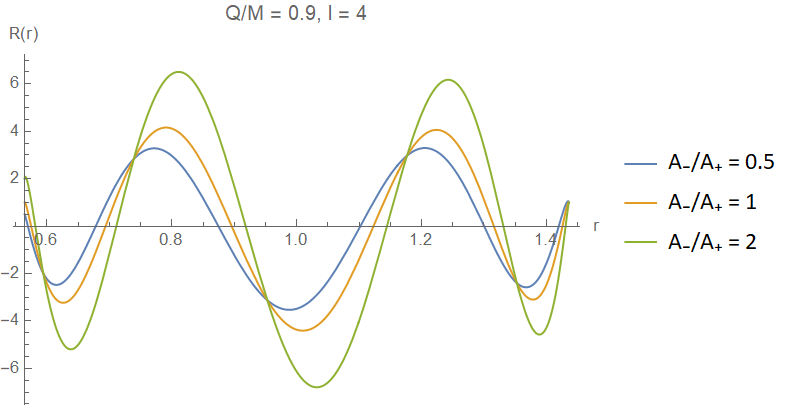}
\end{tabular}
\caption{This plot shows the solutions for the radial functions for the three different $Q/M$ regimes
and $\gamma$ values, all for $l=4$. The color labels are the same as in figure~\ref{fig-gamma-l2}.}
\label{fig:gamma-l4}
\end{figure}

Figure~\ref{fig:gamma-l4} shows the function $R(r)$ when $l = 4$ for different $\gamma$ values for the three BH
$Q/M$ regimes.
As expected, we can see that for the Schwarszchild case there is practically no distinction between the
solutions as $\gamma$ changes. For the typical RN case, the solutions have the same overall form but with
different amplitudes for different $\gamma$ values. As mentioned earlier, the solutions for extremal case
are oscillatory but are slightly different from each other, although it is evident that they are become
more similar as $l$ increases.

\begin{figure}[h]
\begin{tabular}{ccc}
\includegraphics[width=0.3\textwidth]{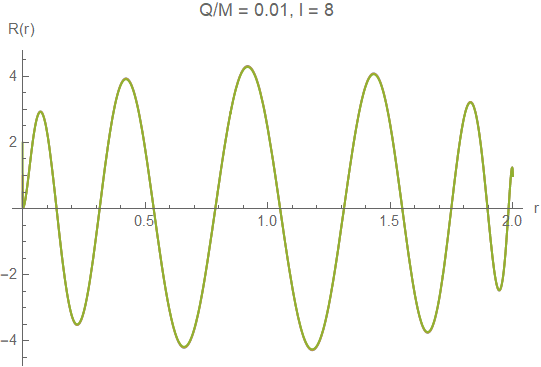} &
\includegraphics[width=0.3\textwidth]{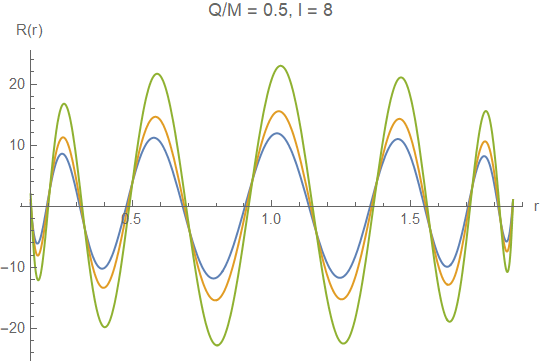} &
\includegraphics[width=0.35\textwidth]{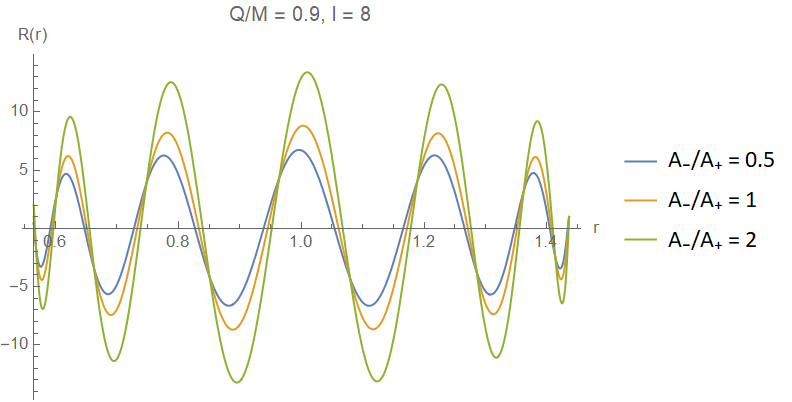}
\end{tabular}
\caption{This plot shows the solutions for the radial functions for the three different $Q/M$ regimes
and $\gamma$ values, all for $l=8$.}
\label{fig:gamma-l8}
\end{figure}

Next, we will explore the plot when we increase $l$ to even higher values. Figure~\ref{fig:gamma-l8}  shows the
solution $R(r)$ for different $\gamma$ values for three different cases of $Q/M$ when we have $l = 8$.

We can see the continuing trend for the Schwarzschild case, and for the other two cases we can see that
the solutions have taken the same form but with varying amplitudes with the highest being for the typical RN
case when $\gamma > 1$.

\begin{figure}[h]
\begin{tabular}{ccc}
\includegraphics[width=0.3\textwidth]{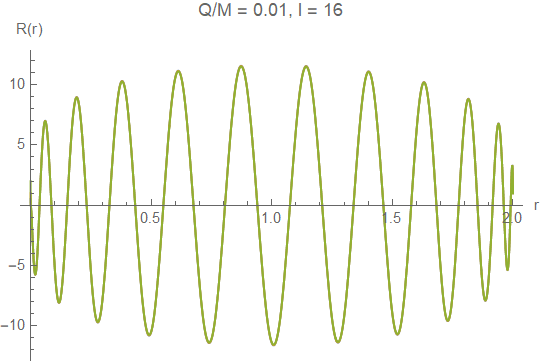} &
\includegraphics[width=0.3\textwidth]{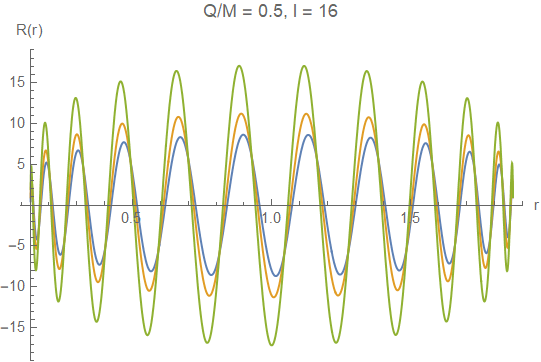} &
\includegraphics[width=0.35\textwidth]{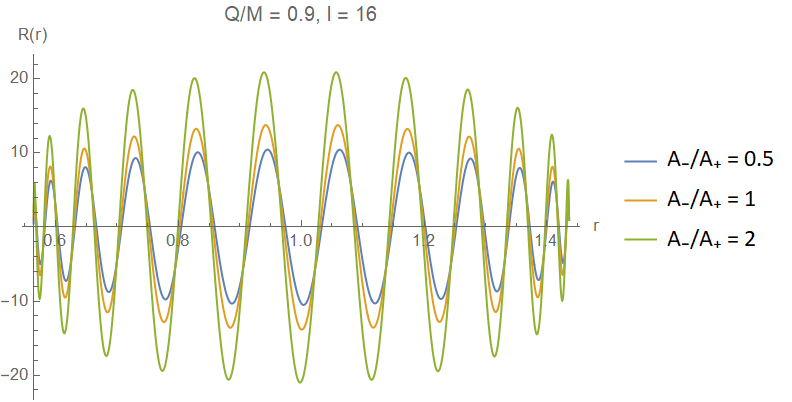}
\end{tabular}
\caption{This plot shows the solutions for the radial functions for the three different $Q/M$ regimes
and $\gamma$ values, all for $l=16$.}
\label{fig:gamma-l16}
\end{figure}

Figure~\ref{fig:gamma-l16} shows the solution $R(r)$ for different $\gamma$ values for three different cases of
$Q/M$ when we have $l = 16$.

The general trend observed is that for nearly neutral BHs ($Q/M = 0.01$), the solutions are
insensitive to $\gamma$ because the inner horizon’s influence is negligible. Moreover, all solutions
become less sensitive to $\gamma$ as $l$ increases; with shorter oscillations distances, it requires
less and less modification to a field profile to shift the value at the outer boundary.
For moderately charged
BHs ($Q/M = 0.5$), the solutions are oscillatory, with the highest amplitudes occurring when
$\gamma = 2$ due to the stronger influence of the inner horizon. For highly-charged BHs
($Q/M = 0.9$), the solutions are oscillatory but with lower amplitudes due to the strong electromagnetic
field and proximity to extremality. The $\gamma = 2$ case still leads to the highest amplitudes.
These results highlight the complex interplay between the charge-to-mass ratio, angular momentum,
and boundary conditions in determining the behavior of perturbations in RN BHs.

\section{Amplitude and $Q/M$ Ratios}
\label{sec-ampl}

Studying the amplitudes of perturbations for different $Q/M$ ratios with altering $l$ values is essential
for understanding the stability, structure, and dynamics of perturbed RN BHs. It provides insights into
energy dissipation mechanisms, tests theoretical models, and helps interpret astrophysical observations.
The dependence of perturbation amplitudes on $Q/M$ and 
$l$ will reveal the complex interplay between gravity and electromagnetism in BH spacetimes, offering a
deeper understanding.

We can expect that for certain values of $l$ there will be one or more particular $Q/M$ values for
which resonance effects can
occur, and the perturbations interact strongly with the BH's electromagnetic field. Throughout this work,
we use the term ``resonance'' in its broader mathematical sense---referring to particularly strong
response in the solution of a differential equation (in this case, meaning the solution of a
boundary value problem) with the parameters of the system take on specific values. The resonance solutions
between the two RN horizons are analogous to standing waves in an optical cavity, which have
steady-state amplitudes that
are enhanced by the quality factor when the wavelength is in resonance with the dimensions of the cavity.
This meaning of ``resonance'' is to be
distinguished from dynamical resonance phenomena involving time-dependent driving forces, which lie
beyond the scope of our stationary perturbation analysis.

\begin{figure}[h]
    \centering
    \includegraphics[width=0.45\linewidth]{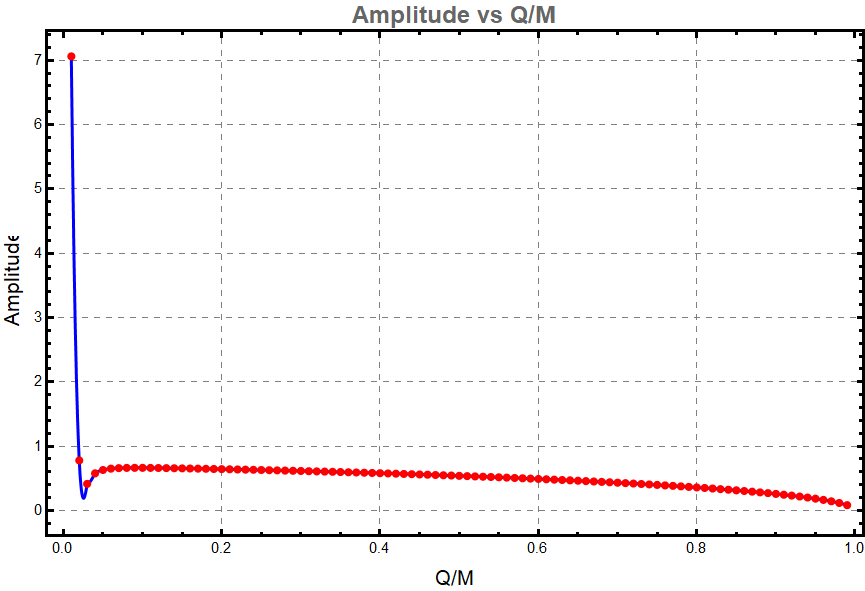}
    \caption{This plot shows the maximum amplitude of $R(r)$ as a function of $Q/M$ for $l = 1$
    and $\gamma = 1$.}
    \label{fig:amp_l_1}
\end{figure}

The stationary resonances
lead to peaks in the amplitudes of the perturbation solutions, which an provide further insights into the
underlying physics. So
in this section we shall characterize and plot the the amplitudes of the solutions for $R(r)$
for a varying set of
$Q/M$ ratios, ranging from $0.01$ to $0.99$.
Figure~\ref{fig:amp_l_1} shows the results for $l=1$. (In Figure~\ref{fig:amp_l_1} and the other
graphs in this section, the plotted amplitudes represent the maximal values of the perturbation functions
in the $r_{-}<r<r_{+}$ region, when the functions are normalized to have $A_{-}=1$. 
Since in most cases the plots are for the case of $\gamma=1$, the normalization is actually
$A_{-}=A_{+}=1$.)

The graph shows that the amplitude is highest for nearly neutral BHs ($Q/M \sim 0$) and
decreases significantly, stabilizing at relatively low values for highly charged BHs ($Q/M \sim 1$).
This behavior is due to the suppression of perturbations by the strong electromagnetic field and the influence
of the inner horizon. Dipole perturbations ($l = 1$) are the lowest-order multipole perturbations and are
less sensitive to the detailed structure of the BH’s geometry and electromagnetic field compared to
higher $l$ values. For $l = 1$, the perturbations are primarily influenced by the overall gravitational
and electromagnetic fields, rather than finer details of the spacetime structure.
As a result, the amplitude decreases smoothly and stabilizes at high $Q/M$, as the electromagnetic field
uniformly suppresses the perturbations. The smooth decrease and stabilization of the amplitude reflect the
simpler nature of dipole perturbations and highlight the stabilizing effect of charge.

\begin{figure}[h]
    \centering
    \includegraphics[width=0.45\linewidth]{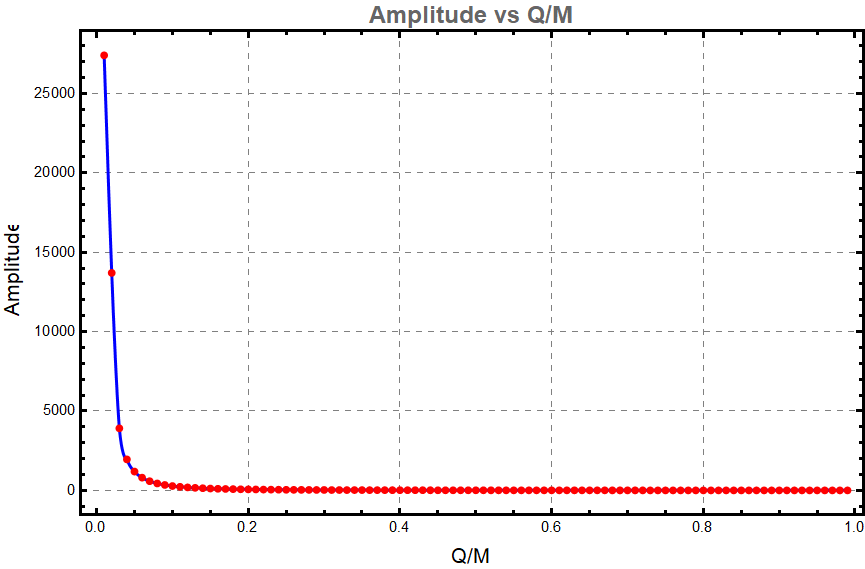}
    \caption{This plot shows the maximum amplitude of $R(r)$ as a function of $Q/M$ for $l = 2$
    and $\gamma = 1$.}
    \label{fig:amp_l_2}
\end{figure}

Figure~\ref{fig:amp_l_2} shows the plot of Amplitude vs. $Q/M$ for $l = 2$.
The graph for quadrupole perturbations($l = 2$) follows a trend very similar to that observed for $l = 1$.
The amplitude is highest when $Q/M \sim 0$ (nearly neutral BH) and decreases significantly as $Q/M$
increases, stabilizing at a low value for highly charged BHs ($Q/M \sim 1$). This similarity
indicates that, like dipole perturbations, quadrupole perturbations are also strongly suppressed by
the presence of a significant electric field as the BH becomes more charged.

\begin{figure}[h]
    \centering
    \includegraphics[width=0.45\linewidth]{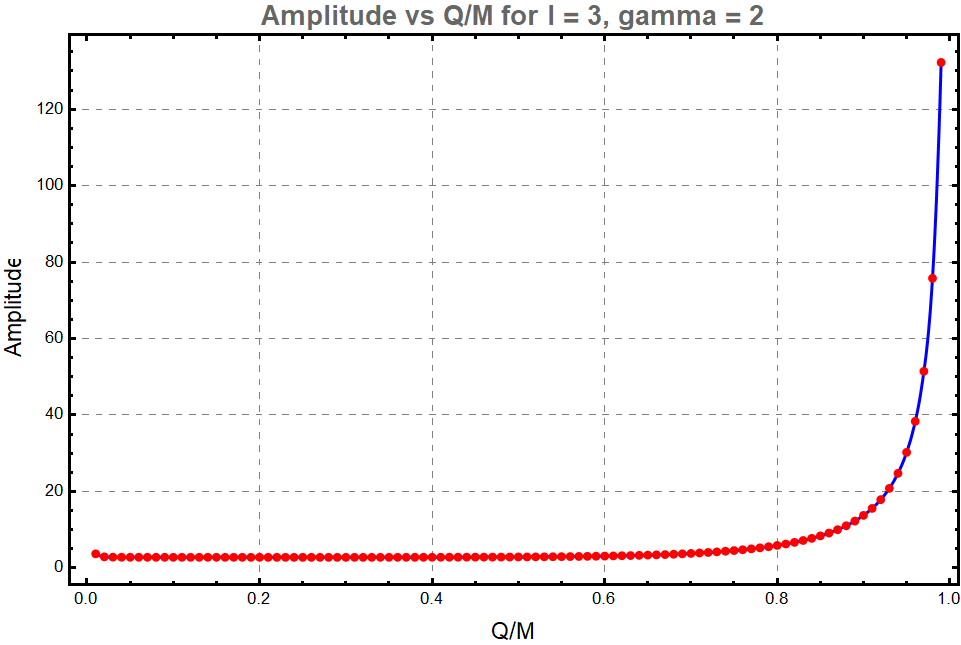}
    \caption{This plot shows the maximum amplitude of $R(r)$ as a function of $Q/M$ for $l = 3$
    and $\gamma = 1$.}
    \label{fig:amp_l_3}
\end{figure}

Figure~\ref{fig:amp_l_3} shows the graph of amplitude versus $Q/M$ for $l = 3$ (octuple perturbations).
We can see a notable difference compared
to the plots for $l = 1$ and $l = 2$. While the amplitude decreases with increasing $Q/M$ for $l = 1$ and
$l = 2$, the amplitude for $l = 3$ increases as $Q/M$ approaches $1$ (highly charged BH).
This opposite trend suggests that octopole perturbations behave differently in the presence of a
strong electric field. The increase in amplitude for $l=3$ may be due to stronger coupling between the
higher multipole moment and the electromagnetic field, leading to enhanced perturbation amplitudes as
the charge-to-mass ratio increases.

\begin{figure}[h]
    \centering
    \includegraphics[width=0.45\linewidth]{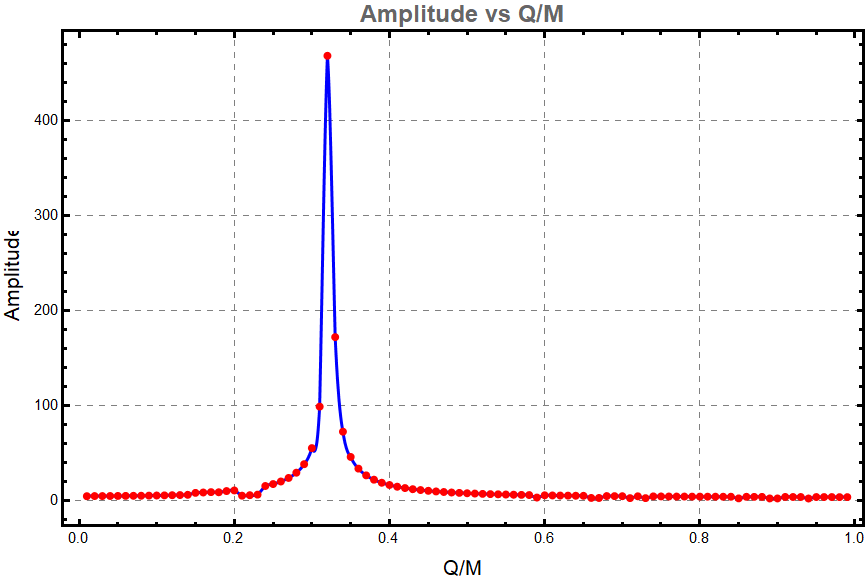}
    \caption{This plot shows the maximum amplitude of $R(r)$ as a function of $Q/M$ for $l = 4$
    and $\gamma = 1$.}
    \label{fig:amp_l_4}
\end{figure}

Figure~\ref{fig:amp_l_4} reveals yet a different kind of behavior: The amplitude reaches a maximum at a
particular value of $Q/M$ and then decreases as $Q/M$ increases further. This behavior can be attributed
to resonance effects and the strong coupling between the hexadecapole perturbations and the BH's
electromagnetic field.

At a specific $Q/M$ value, the wavelength of the perturbations may match with the
electromagnetic field, leading to resonance. This resonance enhances the interaction between the
perturbations and the electromagnetic field, resulting in a peak in the amplitude. However, as
$Q/M$ increases beyond this point, the electromagnetic field becomes even stronger, suppressing the
perturbations and reducing their amplitude. Additionally, the inner horizon $r_-$ becomes more
significant, further contributing to the suppression of the perturbations.

This behavior highlights the importance of the angular momentum parameter $l$ in determining how
perturbations interact with the BH's electromagnetic field. For higher $l$ values, such as $l = 4$, the
perturbations are more sensitive to the detailed structure of the spacetime and electromagnetic field,
leading to more complex behavior, including resonance effects and amplitude peaks at specific $Q/M$ values.

The amplitude peaks observed at specific $Q/M$ values arise from a constructive interference condition
in the inter-horizon region, rather than dynamical resonance. As shown by the WKB analysis in
Appendix~\ref{app:wkb}, when the effective wavenumber $k(r)$ satisfies a quantization condition the
perturbation exhibits maximal amplification, due to phase matching between the
solutions interpolated from the inner and outer boundaries.
The quantization
condition depends critically on $Q/M$, through the horizon locations $r_\pm$ and the function $f(r)$.
For $l=4$, this condition is satisfied at $Q/M\sim 0.3$, explaining the pronounced peak (located
well away from either endpoint) in
Figure~\ref{fig:amp_l_4}. The stationary nature of our analysis means this represents a spatial
resonance---a standing wave pattern optimized by the geometry---rather than a time-dependent
resonant growth.

\begin{figure}[h]
    \centering
    \begin{subfigure}{0.45\textwidth}
        \centering
        \includegraphics[width=\linewidth]{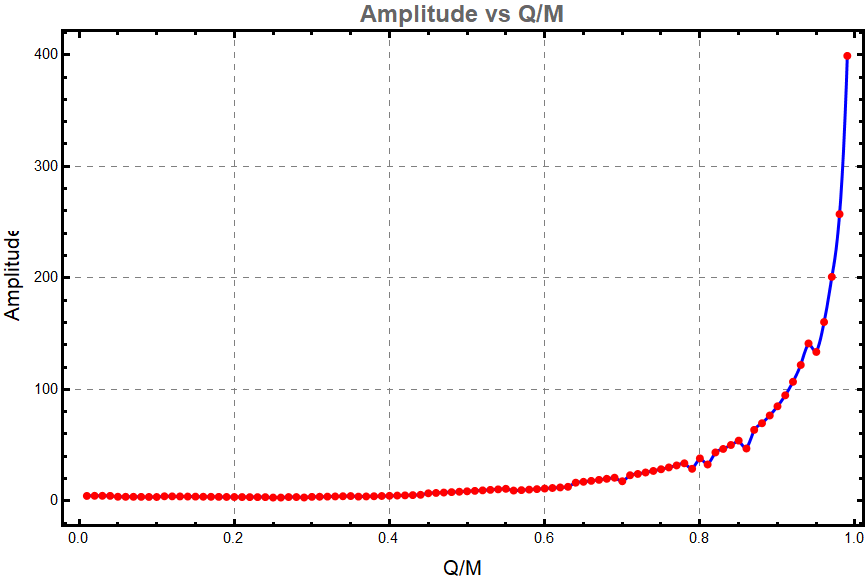}
        \caption{$l = 5$,$\gamma = 1.5$}
        \label{fig:amp_l_5}
    \end{subfigure}
    \hfill
    \begin{subfigure}{0.45\textwidth}
        \centering
        \includegraphics[width=\linewidth]{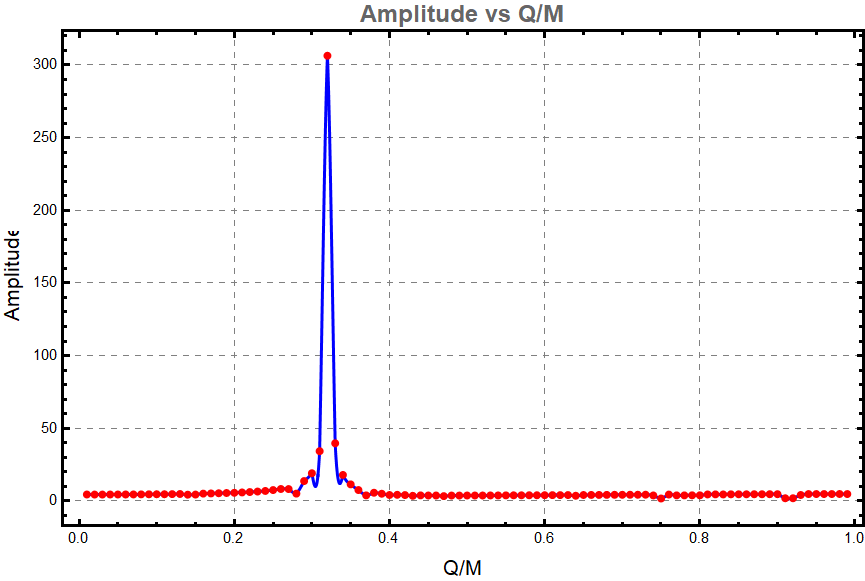}
        \caption{$l = 6$,$\gamma = 1$}
        \label{fig:amp_l_6}
    \end{subfigure}
    
    \vspace{0.5cm}
    
    \begin{subfigure}{0.45\textwidth}
        \centering
        \includegraphics[width=\linewidth]{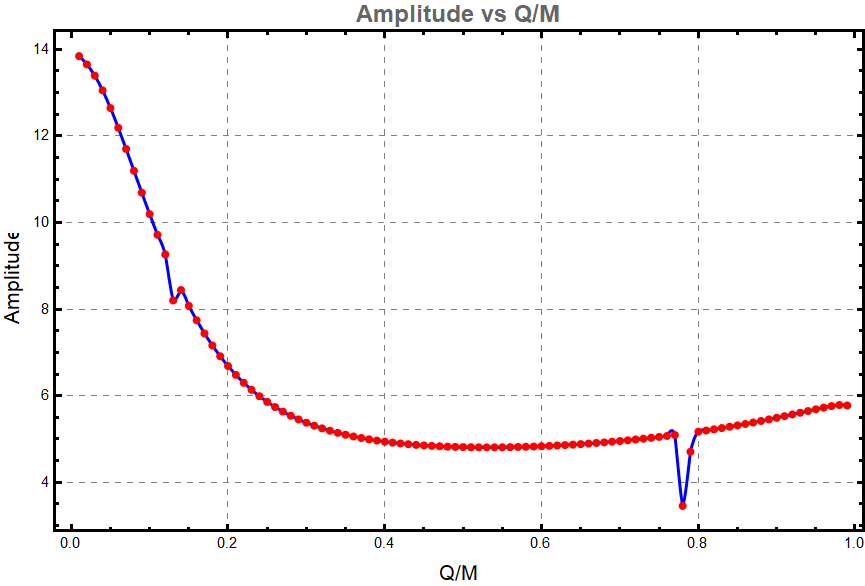}
        \caption{$l = 7$,$\gamma = 1.3$}
        \label{fig:amp_l_7}
    \end{subfigure}
    \hfill
    \begin{subfigure}{0.45\textwidth}
        \centering
        \includegraphics[width=\linewidth]{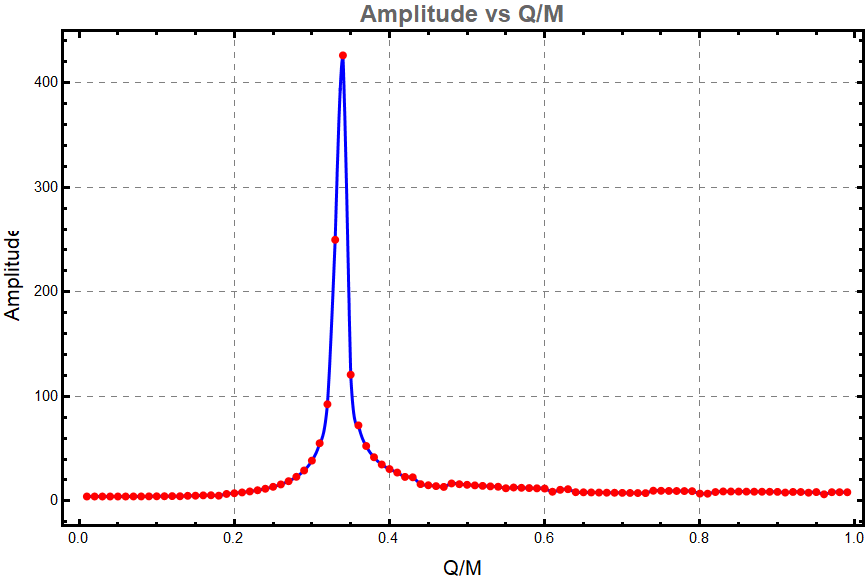}
        \caption{$l = 8$,$\gamma = 1$}
        \label{fig:amp_l_8}
    \end{subfigure}

    \caption{These plots show the amplitude versus $Q/M$ for four higher $l$ values and a variety of
    $\gamma$.}
    \label{fig:amp_l_all}
\end{figure}

The trends observed in the amplitude versus $Q/M$ results for $l = 5$--$8$ (Figure~\ref{fig:amp_l_all})
align closely
with some of the behaviors seen for lower $l$ values. These trends can be grouped into three categories.
We see decreasing amplitude with $Q/M$ for $l = 1$, $2$, and $7$. On the other hand, there is increasing
amplitude with $Q/M$ for $l = 3$ and $l = 5$, at least for the $\gamma$ values in the figures. For
$l = 6$, and $8$, there is a peak in the amplitude at a specific $Q/M$, where resonance effects again cause
the amplitude to rise sharply before being suppressed again at higher $Q/M$. Finally, for $l=7$, a new
phenomenon emerges: pronounced spatial antiresonance at a particular $Q/M$ value where the amplitude is
minimized.

\begin{figure}
    \centering
    \begin{subfigure}{0.45\textwidth}
        \centering
        \includegraphics[width=\linewidth]{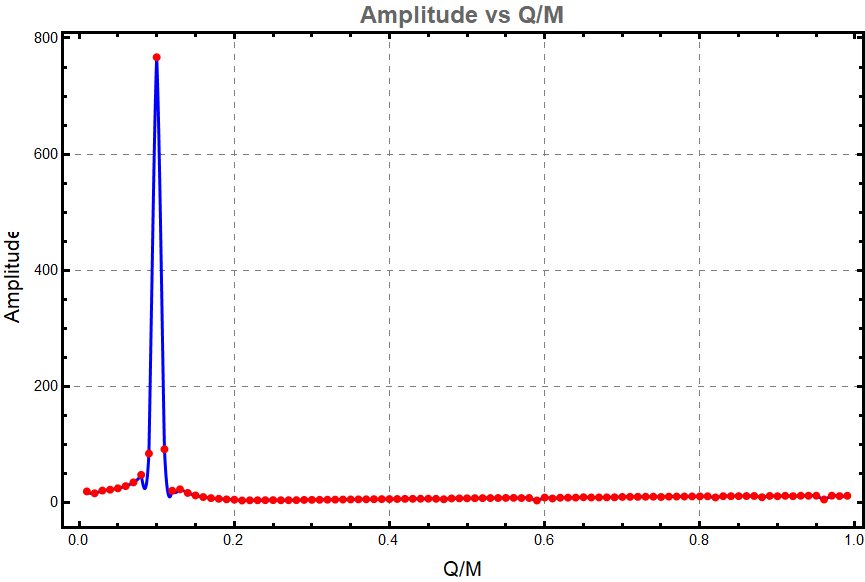}
        \caption{$l = 9$}
        \label{fig:amp_l_9}
    \end{subfigure}
    \hfill
    \begin{subfigure}{0.45\textwidth}
        \centering
        \includegraphics[width=\linewidth]{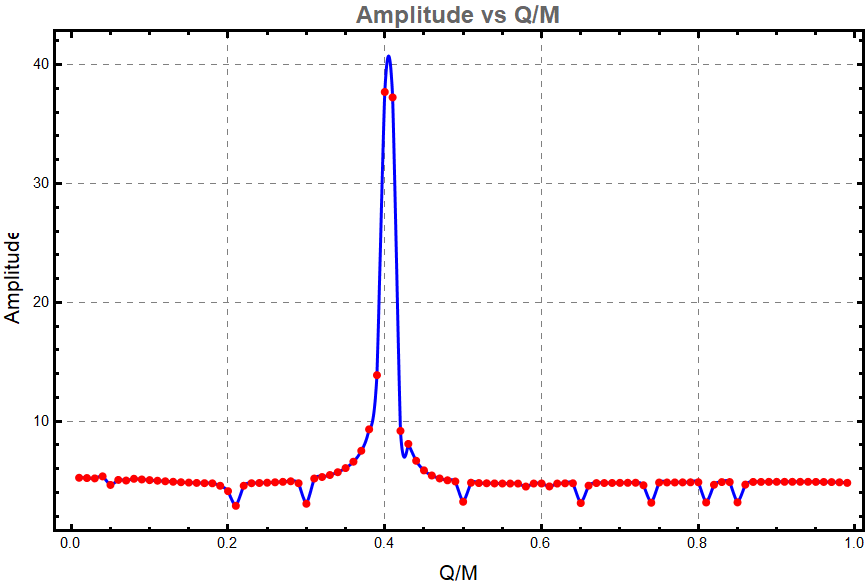}
        \caption{$l = 10$}
        \label{fig:amp_l_10}
    \end{subfigure}
    \hfill
    \begin{subfigure}{0.45\textwidth}
        \centering
        \includegraphics[width=\linewidth]{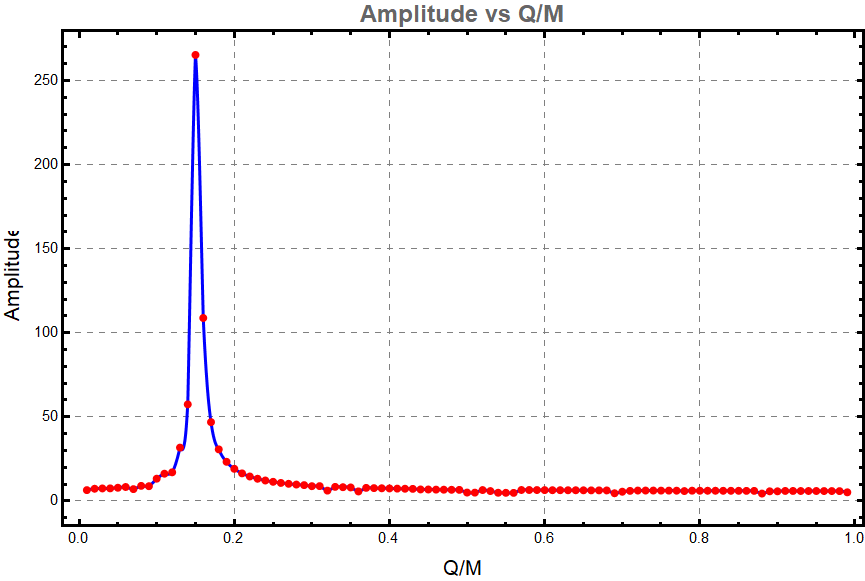}
        \caption{$l = 11$}
        \label{fig:amp_l_11}
    \end{subfigure}
    \hfill
    \begin{subfigure}{0.45\textwidth}
        \centering
        \includegraphics[width=\linewidth]{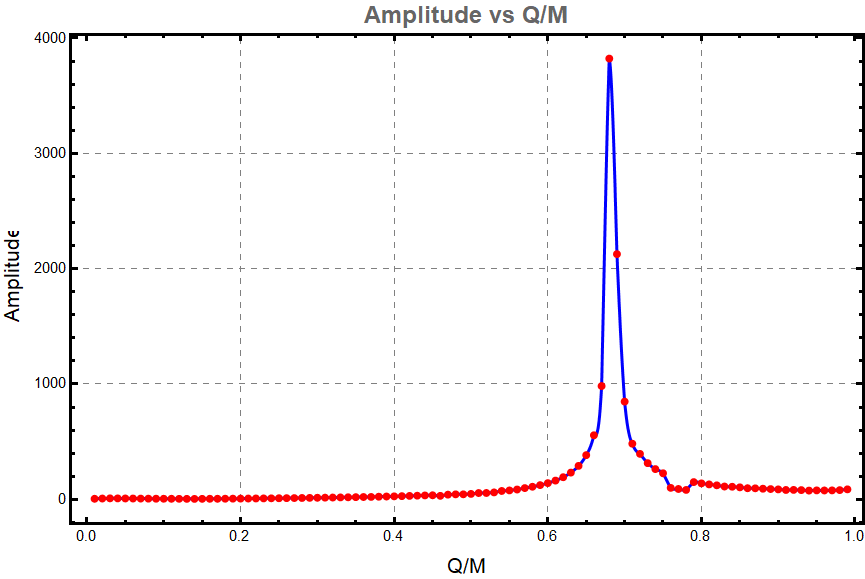}
        \caption{$l = 12$}
        \label{fig:amp_l_12}
    \end{subfigure}

    \vspace{0.5cm}

    \begin{subfigure}{0.45\textwidth}
        \centering
        \includegraphics[width=\linewidth]{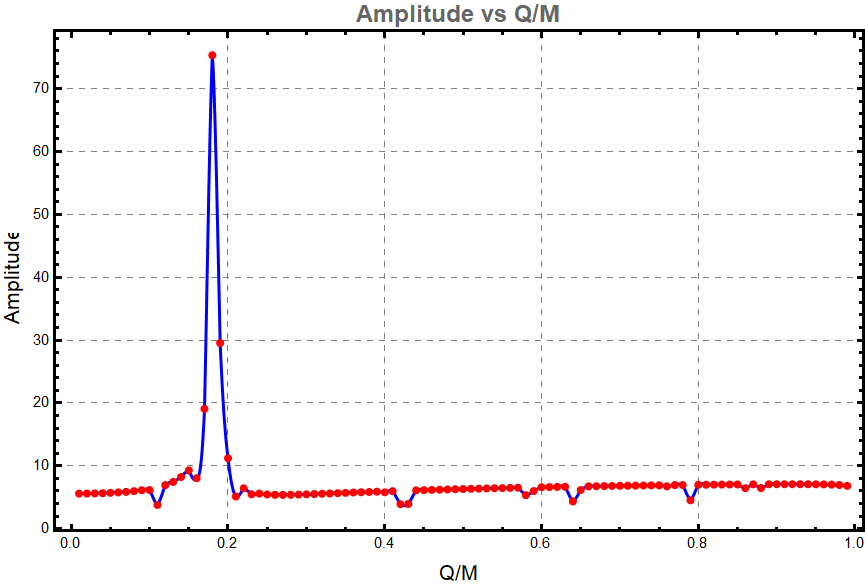}
        \caption{$l = 13$}
        \label{fig:amp_l_13}
    \end{subfigure}
    \hfill
    \begin{subfigure}{0.45\textwidth}
        \centering
        \includegraphics[width=\linewidth]{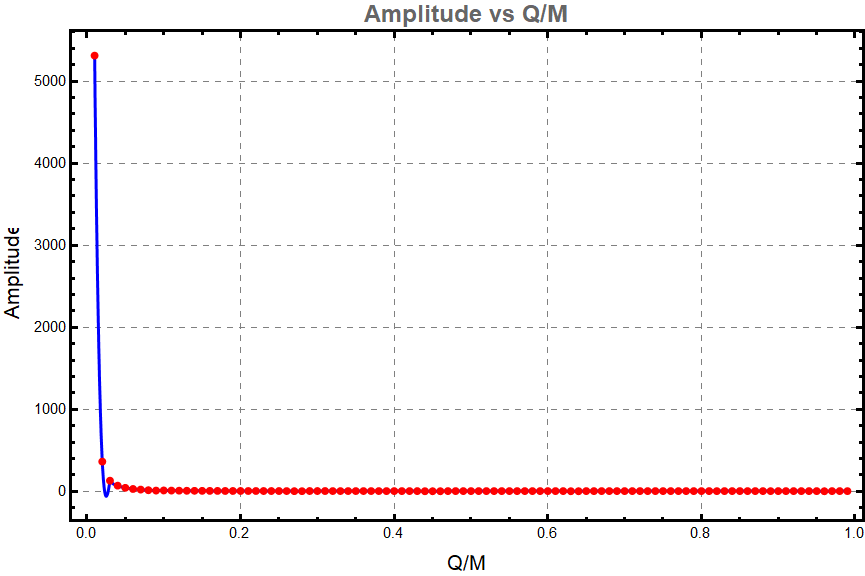}
        \caption{$l = 14$}
        \label{fig:amp_l_14}
    \end{subfigure}
    \hfill
    \begin{subfigure}{0.45\textwidth}
        \centering
        \includegraphics[width=\linewidth]{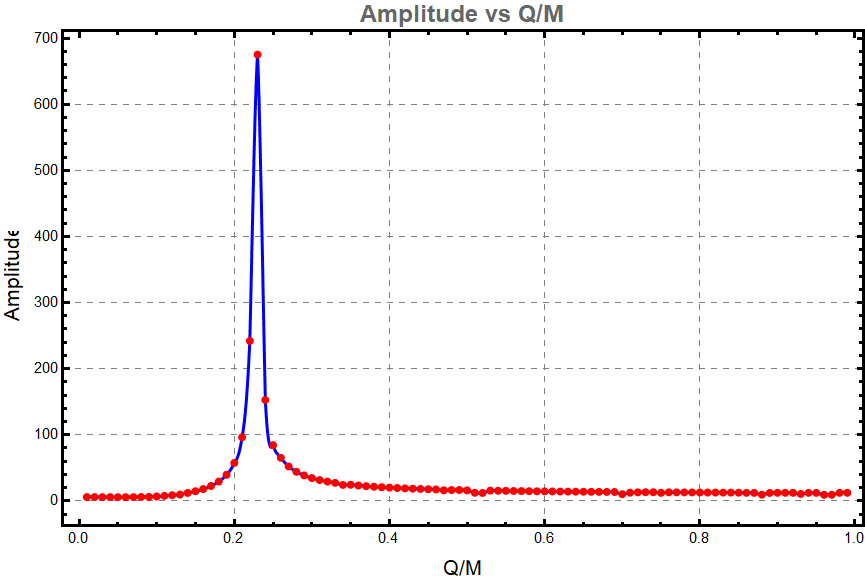}
        \caption{$l = 15$}
        \label{fig:amp_l_15}
    \end{subfigure}
    \hfill
    \begin{subfigure}{0.45\textwidth}
        \centering
        \includegraphics[width=\linewidth]{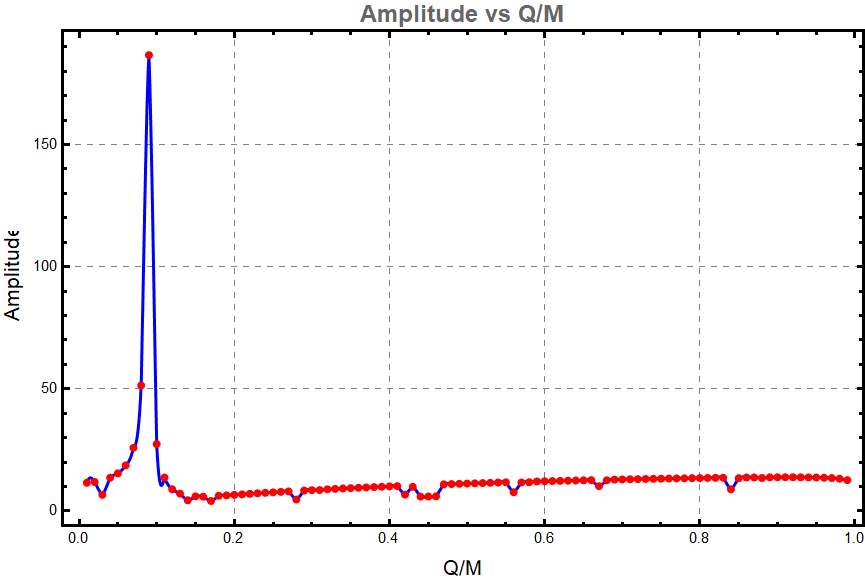}
        \caption{$l = 16$}
        \label{fig:amp_l_16}
    \end{subfigure}

    \caption{These plots show the amplitude versus $Q/M$ for higher $l$ values of $9$--$16$.}
    \label{fig:amp_all_2}
\end{figure}

For even higher $l$ values, which are more sensitive to the detailed structure of the BH's geometry and
electromagnetic field, the strong coupling between the perturbations and the electromagnetic field
enhances the interaction, leading to widespread resonance effects and peak amplitudes at specific $Q/M$
values. This is shown in Figure~\ref{fig:amp_all_2}.
Higher $l$ perturbations probe smaller scales of the spacetime structure, making them more sensitive
to the influence of the horizons and the electromagnetic field. This sensitivity leads to more pronounced
resonance-effects and complicated behavior in the amplitude of the perturbations.

The resonance-like behavior may indicate that energy is efficiently transferred between the perturbation
and the electromagnetic field at specific $Q/M$ values.
Beyond the resonance point, energy would then be dissipated as the perturbations interact with the
strong electromagnetic field and the inner horizon. In astrophysical scenarios, higher $l$ perturbations
may be excited by external disturbances, such as in-falling matter or interactions with other BHs. For lower
$l$ values (such as $l = 1$--$3$), the amplitude trends are simpler, with some cases showing a
monotonic decrease or increase with $Q/M$. For higher $l$ values ($l = 9$--$16$), the resonance-like
behavior becomes more pronounced, reflecting the stronger coupling and increased sensitivity of higher
multipole structures. At some of the $l$ values (especially l=10), there are also multiple antiresonant
$Q/M$ values.

\section{Associated Scalar Field $\Theta$}
\label{sec-Theta}

We shall now return to some secondary theoretical topics. We have already concluded that, for the
frame-dragging $h_{t\phi}$ metric perturbations, we expect there to be no contribution to the Cotton
tensor, because of disagreements in the behavior with respect to time reversal transformations. We
shall now outline how explicit calculations bear this conclusion out.

In CS-modified gravity, the Cotton tensor $C_{\mu\nu}$ introduces parity-violating terms
coupled to the scalar field $\Theta$ through its derivative $v_{\nu} = \partial_{\nu}\Theta$.
For the static, spherically-symmetric RN background, we may decompose
the tensor into two parts and linearize it to first order in perturbations. Of the two parts of the
Cotton tensor, the first is associated with the symmetries of the Ricci curvature tensor, and the second
depends on the dual of the Riemann tensor. These are discussed in more detail in appendix~\ref{app:perturbations}.
In terms of the various curvature tensor components, the two parts of the Cotton tensor are
\begin{equation}\label{eq60}
    \begin{aligned}
        C^{\mu\nu}_{(1)} &= v_{\alpha}\left(\epsilon^{\alpha\mu\sigma\tau}\nabla_{\sigma}R^{\nu}_{\tau}
        + \epsilon^{\alpha\nu\sigma\tau}\nabla_{\sigma}R^{\mu}_{\tau}\right)\\
        C^{\mu\nu}_{(2)} &= v_{\sigma\tau}\left(^*R^{\tau\mu\sigma\nu} + ^*R^{\tau\nu\sigma\mu}\right),
    \end{aligned}
\end{equation}
where we have used $v_{\sigma\tau} = \nabla_{\sigma}v_{\tau}$, which reduces to
$v_{\mu\nu} = -\Gamma^{\alpha}_{\mu\nu} v_{\alpha}$ when the coordinate derivatives of $v_{\nu}$ vanish.

Calculations of the first-order perturbations in these components of the Cotton tensor are shown in
Appendix~\ref{app:perturbations}. The field equations for the perturbation terms are
\begin{equation}
    \delta G_{\mu\nu} + \delta C_{\mu\nu} = 8\pi \delta T_{\mu\nu},
\end{equation}
with the $\delta$ denoting the first-order perturbation part of each quantity.
From the other calculated Riemann tensor components, we get Einstein gravitational field tensor,
\begin{equation}
    G_{tt} = \bar{G}_{tt}, \hspace{0.4cm} G_{ii} = \bar{G}_{ii}, \hspace{0.4cm} G_{ij} = 0
\end{equation}
where, as elsewhere, $\bar{G}_{\mu\nu}$ denotes the background unperturbed RN tensor elements, and $i$ is
a particular spatial index (not summed over).
These relations ensure that $\delta G_{\mu\nu} = 0$, except for $G_{t\phi}$.
Similarly, we can show that for the stress-energy tensor, we have 
\begin{equation}
    T_{tt} = 0, \hspace{0.4cm} T_{ii} = 0, \hspace{0.4cm} T_{ij} = 0
\end{equation}
That makes $\delta T_{\mu\nu} = 0$ (again except for $T_{t\phi}$). So up to the first order in
an axial perturbation, the field equations governing $\Theta$ and its derivatives are mainly of the forms,
\begin{equation}
   \delta C_{tt} = 0, \hspace{0.4cm} \delta C_{ti} = 0, \hspace{0.4cm} \delta C_{ii} = 0, \hspace{0.4cm}
   \delta C_{ij} = 0,
\end{equation}
with the exception of the equation for $C_{t\phi}$.

Up to the first order, we calculated the perturbations in the Cotton tensor to take forms like
\begin{equation}
\begin{aligned}
    &\delta C_{tt} = v_{r}\left(\text{Riemann terms including $H$......}\right) \\
    &+ v_{\theta}\left(\text{Riemann terms including $H$.....}\right) = 0.
    \label{eq-delta-Ctt}
\end{aligned}
\end{equation}
The Riemann terms mentioned are combinations of $H$ and derivatives of $H$ like $\partial_r H$,
$\partial_{\theta} H$, $\partial^2_{r}H$, $\partial^2_{\theta}H$, $\partial_r\partial_{\theta}H$ and so on.
Besides (\ref{eq-delta-Ctt}), the field equations involving the other vanishing Cotton tensor components
are
\begin{equation}
    \begin{aligned}
        \delta C_{tr} &= v_{t}\left(\text{R terms}\ldots\right) = 0\\
        \delta C_{t\theta} &= v_{t}\left(\text{R terms}\ldots\right) = 0\\
        \delta C_{rr} &= v_{r}\left(\text{R terms}\ldots\right)
        + v_{\theta}\left(\text{R terms}\ldots\right) = 0\\
        \delta C_{r\theta} &= v_{r}\left(\text{R terms}\ldots\right)
        + v_{\theta}\left(\text{R terms}\ldots\right) = 0\\
        \delta C_{r\phi} &= v_{\phi}\left(\text{R terms}\ldots\right) = 0\\
        \delta C_{\theta\theta} &= v_{r}\left(\text{R terms}\ldots\right) = 0\\
        \delta C_{\theta\phi} &=  v_{\phi}\left(\text{R terms}\ldots\right) = 0\\
        \delta C_{\phi\phi} &= v_{r}\left(\text{R terms}\ldots\right)
        + v_{\theta}\left(\text{R terms}\ldots\right) = 0.
    \end{aligned}
\end{equation}
As we can see, the vanishing of the expressions for $C_{tr}$ and $C_{t\theta}$ demands that
$\Theta \neq \Theta(t)$---that the field has to be stationary. In the same way, we can see from the
equations for $C_{r\phi}$ and $C_{\theta\phi}$ that $\Theta \neq \Theta(\phi)$; the field must be
dependent of the azimuth. This appears to be physically sound, because of the fact that the
perturbed RN metric's residual
Killing symmetries ($\partial_t$ and  $\partial_\phi$) imply $\Theta$ cannot depend on $t$ or $\phi$.

Moreover, from the equation for $C_{\theta\theta}$, we can see that the trivial solution demands that
$\partial_r{\Theta} = 0$ and hence, the field has to be independent of $r$ as well.
Inserting all these conditions into the equations of $C_{rr}$, $C_{r\theta}$, and $C_{\phi\phi}$,
we can finally see that the vanishing of these terms demands that $\partial_{\theta}\Theta = 0$,
which makes $\Theta$ simply a constant scalar field. The only consistent solution for $\Theta$ in
this perturbative framework is a constant field, with the CS effects actually vanishing---unless
additional symmetry-breaking terms (such as rotation- or time-dependent sources) are introduced. In this
case, the gravitational CS term in the action becomes just a total derivative, proportional to
$^{\star}\mathcal{RR}$, which leads to no independent dynamics. This accords
with (\ref{eq-H-PDE}), which does not depend on $v_{\nu}$ in any way, and with the well-known fact that
there are no parity-odd observables in the RN spacetime with nonvanishing expectations that could be
contracted with $v_{\nu}$.

\section{WKB Approximation}
\label{sec-WKB}

Another potential useful 
In the high-angular-momentum regime $(l \gg 1)$, the radial perturbation equation,
\begin{equation}
\frac{d^2R}{dr^2} + \left(\frac{4M}{f r^3} - \frac{2l(l+1)}{f r^2}\right)R = 0,
\end{equation}
becomes highly oscillatory. This motivates the use of the WKB approximation.
As discussed in appendix B, the general WKB solution takes the form
\begin{equation}
R(r) \sim \frac{1}{\sqrt{k(r)}} \cos\!\left( \int k(r) \, dr + \varphi \right), 
\end{equation}
with effective wavenumber
\begin{equation}
k^2(r) = \frac{4M}{f r^3} - \frac{2l(l+1)}{f r^2}.
\end{equation}
Quantization---meaning a condition that only certain radial profiles are strongly favored over
others between the inner and
outer horizons---follows from the hard wall Bohr–Sommerfeld condition,
\begin{equation}
\int_{r_-}^{r_+} k(r)\,dr = \left(n+1\right)\pi,
\end{equation}
for non-negative integers $n$.

Evaluating this integral (see appendix B), we obtain the analytic WKB quantization rule,
\begin{equation}
\frac{l M \sqrt{2}}{\sqrt{M^2 - Q^2}} = n + 1.
\end{equation}
This relation shows that the number of allowed modes decreases as $Q/M$ increases, with the solution
breaking down in the extremal limit $Q \rightarrow M$. Moreover, at most a single $n$ value may approximately
satisfy this, which accords with what we previously found numerically---that there are particular situations
at large $l$ when the amplitude may be very large, because a half-integer number of wavelengths fit neatly
between the boundaries.

\section{Symmetry in the Extremal Case}
\label{sec-AdS}

The last theoretical topic upon which we have some remarks to make concerns the near-horizon geometry
in the case of an extremal RN metric, for which the charge-to-mass ratio $Q/M \approx 1$.
The behavior of perturbations in the vicinities of BHs can provide critical insights into the stability and
dynamics of BH-dominated spacetime regions. For the extremal case, there is an additional intersting
symmetry. We can show that this symmetry arises due to the near-horizon geometry of the
extremal BH, which exhibits an AdS$_2 \times S^2$ structure. This structure introduces a
conformal symmetry, leading to symmetric solutions for the perturbations.

In the extremal limit, $Q/M = 1$, the BH's horizons becomes degenerate, $r_{-}=r_{+}$,
and the near-horizon line element takes the form:
\begin{equation}
    ds^2 \approx r^2 \, dt^2 - \frac{dr^2}{r^2} - r_+^2 \, d\Omega^2,
\end{equation}
where $r_+$ is the unique horizon radius. This metric is that of the space AdS$_2 \times S^2$,
where AdS$_2$ is two-dimensional Anti-de Sitter space~\cite{ref-dirac1}
(spanned by the coordinates $t$ and $r$), while
the sphere $S^{2}$ is spanned by the angular coordinates.

Anti-de Sitter space is a maximally symmetric spacetime with constant negative curvature. The
AdS$_{2}$ factor in the near-horizon geometry of the extremal RN BH has an isometry group that
includes time translations, dilations (scaling transformations), and special conformal transformations.
The AdS$_{2}$ factor
arises because the radial and time directions near the horizon decouple from the angular directions,
leading to a geometry that is local in nature. This is a general feature of extremal BHs, not just
RN ones. For example, extremal Kerr BHs also exhibit a similar near-horizon geometry with an AdS$_{2}$
factor~\cite{ref-castro,ref-hui}.

These symmetries form the conformal group in $1+1$ dimensions, locally isomorphic to $SO(2,1)$. [Since only
the local---as opposed to global---geometry has the AdS$_{2}$ factor, only local symmetry transformations are
relevant. So while the conformal symmetry of a global AdS$_{2}$ manifold is $SL(2,\mathbb{R})$, which is a
double cover of $SO(2,1)$, there is no need to distinguish between $SL(2,\mathbb{R})$ and $SO(2,1)$
locally, where they are represented by their identical Lie algebras.]
The presence of this conformal symmetry is crucial for understanding the behavior of perturbations in
the extremal limit.

In the extremal limit ($Q/M \approx 1$), the near-horizon geometry dominates
the behavior of the perturbations. The conformal symmetry of AdS$_{2}$ ensures that the solutions are
scale-invariant and symmetric. This explains why the perturbations become symmetric when $Q/M \approx 1$,
as observed in the solutions for different values of $l$.

This reflects the enhanced symmetry of the near-horizon geometry in the extremal limit. It provides
insights into the stability of extremal BHs and their QNMs. It plays a key role
in the AdS/CFT correspondence~\cite{ref-breitenlohner,ref-burgess1,ref-fradkin,ref-aharony},
where the near-horizon geometry of an extremal BH is dual to a
conformal field theory (CFT) in one lower dimension~\cite{ref-mack}.
The symmetry observed in the perturbations of the RN metric in the extremal limit is a direct consequence
of the highly-symmetric AdS$_2 \times S^2$ near-horizon geometry. The conformal symmetry of AdS$_2$
leads to scale-invariant and symmetric solutions, which are a hallmark of extremal BHs.
This symmetry has profound implications for BH physics, holography, and the study of perturbations.

\section{Future Observational Signatures}
\label{sec:observations}

The theoretical framework and results presented in this work, while focused on stationary perturbations,
have implications for future gravitational wave astronomy and tests of fundamental physics. Although
astrophysical BHs are expected to be nearly neutral, even small deviations from neutrality could
leave measurable imprints on gravitational wave signals. We have seen that the specific $h_{t\phi}$
perturbation considered here does not source a CS field directly. However, the qualitative behaviors
we have uncovered---particularly the amplitude suppression, resonance effects, and symmetry enhancement
in the extremal limit---are robust features of parity-violating that should manifest themselves in the
dynamical ringdown signals of charged BHs, both in standard GR and in modified theories. This
section outlines potential observational avenues for probing these effects.

\subsection{Imprints on Gravitational Wave Ringdown}

The ringdown phase of a binary BH merger is dominated by a superposition of the remnant BH's QNMs, which were
dynamically excited by the gravitational interactions between the previous infalling partners. The QNMs'
frequencies and damping rates are determined by the final BH's intrinsic parameters---its mass, spin,
and charge. Our analysis of
the stationary problem can provide a foundation for understanding the more complicated dynamical case.

There are a number of ways in which a parity-violating modification to GR of the type we have considered
might manifest itself. One is the charge-dependent mode suppression our results
show for low $l$ values.
Increasing the charge-to-mass ratio $Q/M$ significantly suppresses the amplitude of low-$l$
$h_{t\phi}$ axial perturbations, as shown in Figures \ref{fig:amp_l_1} and \ref{fig:amp_l_2}).
In a dynamical context, this suggests that for a BH with a non-negligible $Q$, the ringdown signal
in a parity-violating modified gravity model may be intrinsically weaker than for a neutral BH of
the same mass. Future detectors like the Einstein Telescope, Cosmic Explorer, or 
the Laser Interferometer Space Antenna (LISA), with their high sensitivities, could potentially observed
such amplitude deficits. Detecting a ringdown that is systematically quieter than would be expected from a
neutral BH could serve as an indirect indicator of both parity violation and
significant BH charge.

Merger signatures might also be affected by the existence of the spatial resonance modes we have
identified. For perturbations at higher multipoles ($l \geq 4$), we found pronounced amplitude peaks at
specific values of $Q/M$---as in, for example, Figure~\ref{fig:amp_l_4}). The resonance condition arising
from constructive interference of radial modes between the horizons, approximated by the WKB quantization rule
\begin{equation}
\frac{lM\sqrt{2}}{\sqrt{M^{2}-Q^{2}}} = n + 1,
\end{equation}
implies that for a given black hole mass $M$ and charge $Q$, certain higher-angular-overtone modes should be
preferentially excited during a BH's ringdown. The relative amplitudes of different QNMs in a detected
gravitational wave signal could therefore provide a fingerprint for a particular $Q$ value in a modified
gravity theory with parity-violating frame dragging.

The numerical data showing antiresonance for certain $l$ and $Q/M$ values would have similar observable
implications. The obvious difference would be suppression of the ringdown radiation
from the relevant modes, rather than enhancement.

\subsection{Tests with Extreme Mass Ratio Inspirals}

Extreme mass ratio inspirals (EMRIs)---in which a small compact object spirals into a supermassive BH---are
prime targets for future LISA observations. During the inspiral, the small body will effectively act as a
test particle,
probing the spacetime geometry produced by the central supermassive BH. If that central BH carries a charge,
the axial perturbations we have analyzed will be continuously excited in an appropriate modified
gravity theory. The modified frame dragging produced by those perturbations will leave an imprint on the
gravitational radiation emitted by the smaller body as it revolves. 
Our solution for $R(r)$ in the exterior region ($r > r_{+}$), with its characteristic power-law tail
as $r\to\infty$,
\begin{equation}
\label{eq:asymptotic-obs}
R(r) \approx\frac{B_{\infty}}{r^{|\alpha_{-}|}},
\end{equation}
provides an explicit template for comparison to what is to be expected from uncharged BHs or
from RN BHs in unmodified GR.

Moreover, if the infalling body is itself spinning significantly
(as compact objects typically are), a parity-violating $h_{t\phi}$ term will
also generate a Lense–Thirring precession of the falling body's spin. The gravitational wave signature of
the Lense–Thirring effect of an orbiting body is a phenomenon that has not been that extensively studied
at this point. However, it is known that such radiation would exist~\cite{ref-romero}.

\subsection{The Role of the CS Field}

While the stationary $h_{t\phi}$ perturbation forced the CS scalar $\Theta$ to be constant,
this conclusion is particular to the high degree of symmetry we imposed. In a more general
scenario---such as a time-dependent ringdown, a binary inspiral, or a both charged and rotating
Kerr–Newman background metric---the Cotton tensor will definitely not be expected
to vanish. The time-derivative of the perturbation $\dot{h}_{t\phi}$ has the correct discrete
symmetries to be linearly related to a CPT-violating gravitational CS term.

In such cases, the CS field will be dynamically sourced, leading to two key observables, corresponding
to differing behaviors for right- and left-circularly polarized gravitational waves~\cite{ref-altschul44}.
The first is amplitude birefringence, via which the two circular polarizations would be absorbed or damped
at different rates, leading to net ellipticity even an initial gravitational wave signal had none.
The second is velocity birefringence (something which has a familiar analogue in the
propagation of electromagnetic waves in magnetized plasmas and other chiral media), in which
the right- and left-handed
gravitational wave polarizations would propagate with slightly different phase velocities,
causing a frequency-dependent phase difference that can accumulate significantly over cosmological distances.

Future detectors like LISA and the advanced ground-based observatories that will be sensitive to both
polarization modes could search for these birefringence effects in the ringdown signals of black holes,
providing a direct observational probe of the parity-violating sector of gravity. Our results, even in
their null form for this specific case, underscore the necessity of additional symmetry breaking
to unlock these unique CS signatures.

\subsection{Kerr Versus Parity-Violating Frame Dragging}

Both the standard Kerr geometry and the parity violation scenarios we have considered have metrics with
nonvanishing $h_{t\phi}$ components. These time-angle metric components naturally produce
frame dragging between their inner and outer horizons, and some of the observable effects of these
two metric types may be similar. However,
resonance conditions may provide a way of distinguishing observations between these two scenarios.
From the WKB results, we know that the resonance condition depends on $Q/M$ in a nontrivial fashion---and
in a way that is quite different from the dependence of $h_{t\phi}$ on the angular momentum parameter $a$
in the Kerr metric.

\section{Conclusions and Outlook}
\label{sec-concl}

\subsection{Results}

In this work, we have conducted a comprehensive analysis of axial perturbations in RN BH spacetimes,
including possible effects of parity violation, focusing on the interplay between charge, angular
dependence, and boundary conditions. By solving the governing
perturbation equations analytically in the fully interior and exterior
asymptotic regimes and numerically in the bounded region between
the Cauchy and event horizons, we uncovered several key features.
\begin{itemize}
    \item \textbf{Charge-dependent behavior:} The electromagnetic field tends to suppress perturbation
     amplitudes for increasing charge-to-mass ratios $Q/M$, with nearly perfect in-out symmetrization of
     solutions in the extremal limit ($Q/M \sim 1$). This reflects the emergent AdS$_2\times S^{2}$
     near-horizon geometry, whose conformal symmetry constrains the perturbation spectrum.
    \item \textbf{Angular momentum hierarchy:} For low $l$, perturbations exhibit a monotonic amplitude
    decrease or increase with $Q/M$, while higher $l$ modes ($l \geq 4$) show resonance-like peaks at critical
    $Q/M$ values, signaling enhanced coupling between the electric field and higher multipoles.
    \item \textbf{Role of the CS scalar $\Theta$:} Consistency demands $\Theta$ to be constant in this
    framework, unless additional symmetry-breaking terms (e.g., rotation- or time-dependent backgrounds)
    are present. This rigidity underscores the subtle balance between posible
    CS modifications to gravitation and discrete spacetime
    symmetries. While our investigations were originally motivated by interest in CS gravity specifically,
    our perturbation calculations are ultimately valid outside the CS theory.
    \item \textbf{WKB quantization:} High-$l$ modes obey a quantized spectrum tied to $Q/M$,
    corresponding to the resonances in the perturbation amplitude.
    However, the solutions break down in the extremal limit---a hint toward potential non-perturbative effects
    or instabilities.
\end{itemize}

Our results bridge a critical gap in the study of charged BHs in modified gravity. The observed resonance
effects and amplitude suppression could influence gravitational wave signatures from charged compact
objects, offering ideas for observational tests for future detectors such as LISA. Moreover, the constraints
on $\Theta$ highlight the need for broader analyses as mentioned below to fully assess a theory's viability.

\subsection{Future Prospects}
Building on the present work, several theoretical extensions are worth pursuing.
\begin{itemize}
\item \textbf{Nonlinear perturbations:} incorporating higher-order effects to capture back-reaction of
the perturbations on the background metric and to incorporate nonzero CS phenomena.
\item \textbf{Rotating charged BHs:} extending the analysis to Kerr–Newman geometries in CS gravity,
where both rotation and can charge interact with parity- and time-reversal-violating terms,
and there is no expectation that the Cotton tensor might vanish.
\item \textbf{Dynamical CS scalar fields:} allowing $\Theta$ to evolve in time,
thereby testing the robustness of the constant-field result.
\item \textbf{Holographic connections:} exploiting the AdS$_2 \times S^2$ symmetry in the extremal limit
to link perturbation spectra with dual conformal field theories.
BHs that are born in near-extremal states (potentially from neutron star mergers or accretion
processes) could serve as laboratories for testing the conformal behavior involved.
\item \textbf{Experimental tests:} using amplitude modulation, the resonant excitation of high-overtone modes,
and 
to provide observational handles to measure BH charge, test the nature of gravity in strong-field regimes,
and search for parity violation. As gravitational wave astronomy enters its next phase, theoretical
predictions will become increasingly testable,
potentially transforming our results from a purely theoretical
analysis into a blueprint for experimental discovery.
\end{itemize}

Together, these directions promise a deeper theoretical understanding of how CS modifications really could
interact with BH spacetimes, and they provide a natural bridge between classical perturbation theory
and broader frameworks in gravitational physics. By elucidating the rich behavior of perturbations
in this regime, our work advances the theoretical toolkit for testing modified gravity
theories and motivates further exploration of charged BHs as laboratories for fundamental physics.

\appendix

\section{Appendix: Perturbed Field Equations}
\label{app:perturbations}

\subsection{Perturbations to Tensors}

We need to consider a perturbed metric in the form
 \begin{equation}
     g_{\mu\nu} = \bar{g}_{\mu\nu} + \epsilon h_{\mu\nu},
 \end{equation}
where the background metric is not flat; that is, the first derivatives of the background metric with respect
to the coordinates are non-vanishing. For such a general perturbation, the inverse must be of the form
 \begin{equation}
     g^{\mu\nu} = \Bar{g}^{\mu\nu} + \epsilon h^{\mu\nu},
 \end{equation}
Assuming this  holds, then we know that the specific form of the inverse of $h_{\mu\nu}$ can be
written as
 \begin{equation}
      h^{\mu\nu} = -\Bar{g}^{\mu\alpha}h_{\alpha\beta}\Bar{g}^{\nu\beta}.
 \end{equation}
So the full inverse of the metric tensor has the form
\begin{equation}
    g^{\mu\nu} = \Bar{g}^{\mu\nu} - \epsilon \Bar{g}^{\mu\alpha}h_{\alpha\beta}\Bar{g}^{\nu\beta}.
\end{equation}
Using this the Christoffel symbols can be calculated. The general form of an affine connection is
\begin{equation}
    \Gamma^{\lambda}_{\mu\nu} = \frac{1}{2}g^{\lambda\sigma}\left(\partial_{\mu}g_{\nu\sigma} +
    \partial_{\nu}g_{\mu\sigma} - \partial_{\sigma}g_{\mu\nu}\right).
\end{equation}
Expanding the terms up to first order in $\epsilon$, we get
\begin{equation}
    \Gamma^{\lambda}_{\mu\nu} = \Bar{\Gamma}^{\lambda}_{\mu\nu} + \frac{\epsilon}{2}
    \left[\Bar{g}^{\lambda\sigma}\left(\partial_{\nu}h_{\mu\sigma} + \partial_{\mu}h_{\nu\sigma}
    - \partial_{\sigma}h_{\mu\nu}\right) + h^{\lambda\sigma}\left(\partial_{\nu}\Bar{g}_{\mu\sigma}
    + \partial_{\mu}\Bar{g}_{\nu\sigma} - \partial_{\sigma}\Bar{g}_{\mu\nu}\right)\right].
\end{equation}
We make an assumption that the second term on the right-hand side forms a third-rank tensor, denoted
by $P^{\lambda}_{\mu\nu}$. Using this the expression simplifies to
\begin{equation}
    \Gamma^{\lambda}_{\mu\nu} = \Bar{\Gamma}^{\lambda}_{\mu\nu} + \epsilon P^{\lambda}_{\mu\nu},
\end{equation}
where we have again
\begin{equation}
    P^{\lambda}_{\mu\nu} = \frac{1}{2}\left[\Bar{g}^{\lambda\sigma}\left(\partial_{\nu}h_{\mu\sigma}
    + \partial_{\mu}h_{\nu\sigma} - \partial_{\sigma}h_{\mu\nu}\right)
    + h^{\lambda\sigma}\left(\partial_{\nu}\Bar{g}_{\mu\sigma} + \partial_{\mu}\Bar{g}_{\nu\sigma}
    - \partial_{\sigma}\Bar{g}_{\mu\nu}\right)\right].
\end{equation}

Now using this form of the Christoffel symbol we can calculate the Riemann tensor which comes out to be
as follows:
\begin{equation}
    \tensor{R}{^{\lambda}_{\mu\nu\rho}} = \tensor{\Bar{R}}{^{\lambda}_{\mu\nu\rho}}
    + \epsilon\left[\partial_{\nu}P^{\lambda}_{\mu\rho} - \partial_{\rho}P^{\lambda}_{\mu\nu}
    + P^{\lambda}_{\sigma\nu}\Bar{\Gamma}^{\sigma}_{\mu\rho}
    + \Bar{\Gamma}^{\lambda}_{\sigma\nu}P{\Gamma}^{\sigma}_{\mu\rho}
    - P^{\lambda}_{\sigma\rho}\Bar{\Gamma}^{\sigma}_{\mu\nu}
    - \Bar{\Gamma}^{\lambda}_{\sigma\nu}P^{\sigma}_{\mu\rho}\right] + \mathcal{O}(\epsilon^2).
\end{equation}
The perturbation term (the second on the right-hand side) is denoted
$\delta\tensor{R}{^{\lambda}_{\mu\nu\rho}}$. Then the expression for this perturbation in the
Riemann tensor can be isolated as
\begin{equation}\label{eq01}
    \delta\tensor{R}{^{\lambda}_{\mu\nu\rho}} = \partial_{\nu}P^{\lambda}_{\mu\rho}
    - \partial_{\rho}P^{\lambda}_{\mu\nu} + P^{\lambda}_{\sigma\nu}\Bar{\Gamma}^{\sigma}_{\mu\rho}
    + \Bar{\Gamma}^{\lambda}_{\sigma\nu}P^{\sigma}_{\mu\rho}
    - P^{\lambda}_{\sigma\rho}\Bar{\Gamma}^{\sigma}_{\mu\nu}
    - \Bar{\Gamma}^{\lambda}_{\sigma\nu}P^{\sigma}_{\mu\rho}.
\end{equation}
This can be further simplified using the fact that the unperturbed covariant derivative of the third-rank
tensor $P^{\lambda}_{\mu\nu}$ is
\begin{equation}
    \Bar{\nabla}_{\rho}P^{\lambda}_{\mu\nu} = \partial_{\rho}P^{\lambda}_{\mu\nu}
    + \Bar{\Gamma}^{\lambda}_{\rho\sigma}P^{\sigma}_{\mu\nu}
    - \Bar{\Gamma}^{\sigma}_{\mu\rho}P^{\lambda}_{\sigma\nu}
    - \Bar{\Gamma}^{\sigma}_{\nu\rho}P^{\lambda}_{\mu\sigma},
\end{equation}
where $\Bar{\nabla}$ represents the covariant derivative with respect to the background metric
$\Bar{g}_{\mu\nu}$.
Therefore the expression in (\ref{eq01}) can be simplified to
\begin{equation}
    \begin{aligned}
        \delta\tensor{R}{^{\lambda}_{\mu\nu\rho}} &= \partial_{\nu}P^{\lambda}_{\mu\rho} 
        - \partial_{\rho}P^{\lambda}_{\mu\nu} + P^{\lambda}_{\sigma\nu}\Bar{\Gamma}^{\sigma}_{\mu\rho} 
        + \Bar{\Gamma}^{\lambda}_{\sigma\nu}P^{\sigma}_{\mu\rho}
        - P^{\lambda}_{\sigma\rho}\Bar{\Gamma}^{\sigma}_{\mu\nu}
        - \Bar{\Gamma}^{\lambda}_{\sigma\nu}P^{\sigma}_{\mu\rho}\\
        &= \left(\partial_{\nu}P^{\lambda}_{\mu\rho} + \Bar{\Gamma}^{\lambda}_{\sigma\nu}P^{\sigma}_{\mu\rho}
        - \Bar{\Gamma}^{\sigma}_{\rho\nu}P^{\lambda}_{\mu\sigma}
        - \Bar{\Gamma}^{\sigma}_{\mu\nu}P^{\lambda}_{\rho\sigma}\right) -
        \left(\partial_{\rho}P^{\lambda}_{\mu\nu} + \Bar{\Gamma}^{\lambda}_{\sigma\nu}P^{\sigma}_{\mu\rho}
        - \Bar{\Gamma}^{\sigma}_{\rho\nu}P^{\lambda}_{\mu\sigma}
        - \Bar{\Gamma}^{\sigma}_{\mu\rho}P^{\lambda}_{\sigma\nu} \right)\\
        &= \Bar{\nabla}_{\nu}P^{\lambda}_{\mu\rho} - \Bar{\nabla}_{\rho}P^{\lambda}_{\mu\nu}.
    \end{aligned}
\end{equation}
Thus we can write the Riemann tensor in the compact form
\begin{equation}
    \tensor{R}{^{\lambda}_{\mu\nu\rho}} = \tensor{\Bar{R}}{^{\lambda}_{\mu\nu\rho}}
    + \epsilon \delta\tensor{R}{^{\lambda}_{\mu\nu\rho}} + \mathcal{O}(\epsilon^2),
\end{equation}
where the perturbation term is
\begin{equation}
    \delta\tensor{R}{^{\lambda}_{\mu\nu\rho}} = \Bar{\nabla}_{\nu}P^{\lambda}_{\mu\rho}
    - \Bar{\nabla}_{\rho}P^{\lambda}_{\mu\nu}.
    \label{eq-delta-R}
\end{equation}

Using this we can calculate the perturbation in the Ricci tensor by contracting the indices $\lambda$ and
$\nu$,
\begin{equation}
    R_{\mu\rho} = \tensor{R}{^{\lambda}_{\mu\lambda\rho}} = \Bar{R}_{\mu\rho} + \delta R_{\mu\rho},
\end{equation}
where we have
\begin{equation}
    \delta R_{\mu\rho} = \delta\tensor{R}{^{\lambda}_{\mu\lambda\rho}} =
    \Bar{\nabla}_{\lambda}P^{\lambda}_{\mu\rho} - \Bar{\nabla}_{\rho}P^{\lambda}_{\mu\lambda}
\end{equation}
Similarly we can calculate the perturbation to the Ricci scalar as follows:
\begin{equation}
    \begin{aligned}
        R &= g^{\mu\rho}R_{\mu\rho}\\
        &= \left(\Bar{g}^{\mu\rho} + \epsilon h^{\mu\rho}\right)\left(\Bar{R}_{\mu\rho}
        + \delta R_{\mu\rho}\right)\\
        &= \Bar{R} + \epsilon\left(\Bar{g}^{\mu\rho}\delta R_{\mu\rho} + h^{\mu\rho}\Bar{R}_{\mu\rho}\right).
    \end{aligned}
\end{equation}
Hence, the perturbation in the Ricci scalar can be written as
\begin{equation}
    \delta R = \left[\Bar{g}^{\mu\rho}\left(\Bar{\nabla}_{\lambda}P^{\lambda}_{\mu\rho}
    - \Bar{\nabla}_{\rho}P^{\lambda}_{\mu\lambda}\right) + h^{\mu\rho}\Bar{R}_{\mu\rho}\right].
\end{equation}
Making use of all these perturbed forms, we can calculate the first-order perturbation to the Einstein
field tensor. That is
\begin{equation}
    \begin{aligned}
        G_{\mu\nu} &= R_{\mu\nu} - \frac{1}{2}g_{\mu\nu}R\\
        &= \left(\Bar{R}_{\mu\nu} + \delta R_{\mu\nu}\right) - \frac{1}{2}\left(\Bar{g}_{\mu\nu}
        + \epsilon h_{\mu\nu}\right)\left(\Bar{R} + \delta R\right)\\
        &= \Bar{G}_{\mu\nu} + \epsilon\left[\delta R_{\mu\nu} - \frac{1}{2}h_{\mu\nu}\Bar{R}
        - \frac{1}{2}\Bar{g}_{\mu\nu}\left(\Bar{g}^{\alpha\beta}\delta R_{\alpha\beta}
        + h^{\alpha\beta}\Bar{R}_{\alpha\beta}\right)\right] + \mathcal{O}(\epsilon^2).
    \end{aligned}
\end{equation}
The perturbation in the Einstein tensor up to the first order can also be written
\begin{equation}
    \delta G_{\mu\nu} = \delta R_{\mu\nu} - \frac{1}{2}\left(h_{\mu\nu}\Bar{R}
    + \Bar{g}_{\mu\nu}\Bar{g}^{\alpha\beta}\delta R_{\alpha\beta}
    + \Bar{g}_{\mu\nu}h^{\alpha\beta}\Bar{R}_{\alpha\beta}\right).
\end{equation}

In a theory with a Chern-Simmons background, we get an additional second-rank tensor in the field
equations---which
is now commonly known as the Cotton tensor (generalizing the original Cotton tensor, which was only defined
in three dimensions). The Cotton tensor has two parts to it---one which associates itself with
the symmetries of the Ricci tensor and the other which consists of a dual form of the Riemann tensor. We
shall denote them below as the  first and second parts of the Cotton tensor, $C^{\mu\nu}_{(1)}$ and
$C^{\mu\nu}_{(2)}$, respectively. In terms of Riemann tensor components,
we have these expressions for the Cotton tensor terms:
\begin{equation}
    \begin{aligned}
        C^{\mu\nu}_{(1)} &= v_{\alpha}\left(\epsilon^{\alpha\mu\sigma\tau}\nabla_{\sigma}R^{\nu}_{\tau}
        + \epsilon^{\alpha\nu\sigma\tau}\nabla_{\sigma}R^{\mu}_{\tau}\right)\\
        C^{\mu\nu}_{(2)} &= v_{\sigma\tau}\left(^*R^{\tau\mu\sigma\nu} + ^*R^{\tau\nu\sigma\mu}\right).
    \end{aligned}
\end{equation}
These terms involve $v_{\alpha}$, which is the the derivative $\partial_{\alpha}\Theta$ of the Chern-Simons
scalar field (which sets a preferred background direction in spacetime---often taken to be purely spacelike,
so that parity and boost symmetries are broken, but not spatial isotropy) and the further derivative
$v_{\sigma\tau} = \nabla_{\sigma}v_{\tau}$.

Calculating the first-order perturbations to these components of the Cotton tensor can be done as shown
below. We first start with the first part: 
\begin{equation}
    \begin{aligned}
        C^{\mu\nu}_{(1)} &= v_{\alpha}\left(\epsilon^{\alpha\mu\sigma\tau}\nabla_{\sigma}R^{\nu}_{\tau}
        + \epsilon^{\alpha\nu\sigma\tau}\nabla_{\sigma}R^{\mu}_{\tau}\right)\\
        &= v_{\alpha}\left(\epsilon^{\alpha\mu\sigma\tau}\nabla_{\sigma}\left(R^{\nu}_{\tau}
        + \delta R^{\nu}_{\tau}\right) + \epsilon^{\alpha\nu\sigma\tau}\nabla_{\sigma}\left(R^{\mu}_{\tau}
        + \delta R^{\mu}_{\tau}\right)\right)\\
        &= \Bar{C}^{\mu\nu}_{(1)}
        + \epsilon\left[v_{\alpha}\left(\epsilon^{\alpha\mu\sigma\tau}\nabla_{\sigma}\delta R^{\nu}_{\tau}
        + \epsilon^{\alpha\nu\sigma\tau}\nabla_{\sigma}\delta R^{\mu}_{\tau}\right)\right]
    \end{aligned}
\end{equation}
Thus we can write the first part of Cotton tensor as
\begin{equation}
    C^{\mu\nu}_{(1)} = \Bar{C}^{\mu\nu}_{(1)} + \epsilon\delta C^{\mu\nu}_{(1)},
\end{equation}
where the perturbation is
\begin{equation}
    \delta C^{\mu\nu}_{(1)} = \left[v_{\alpha}\left(\epsilon^{\alpha\mu\sigma\tau}\nabla_{\sigma}
    \delta R^{\nu}_{\tau} + \epsilon^{\alpha\nu\sigma\tau}\nabla_{\sigma}\delta R^{\mu}_{\tau}\right)\right],
\end{equation}
which can be also written equivalently with lowered indices,
\begin{equation}
    \delta C^{(1)}_{\mu\nu} = \left[v_{\alpha}\left(\epsilon^{\alpha\beta\sigma\tau}
    \Bar{g}_{\mu\beta}\nabla_{\sigma}\delta R_{\nu\tau}
    + \epsilon^{\alpha\gamma\sigma\tau}\Bar{g}_{\nu\gamma}\nabla_{\sigma}\delta R_{\mu\tau}\right)\right].
\end{equation}
Similarly, for the second part of Cotton tensor we get
\begin{equation}
    \begin{aligned}
        C^{\mu\nu}_{(2)} &= v_{\sigma\tau}\left(^*R^{\tau\mu\sigma\nu}
        + ^*R^{\tau\nu\sigma\mu}\right)\\
        &= v_{\sigma\tau}\left(g^{\mu\lambda}\,^*\tensor{R}{^\tau_\lambda^{\sigma\nu}}
        + g^{\nu\lambda}\, ^*\tensor{R}{^\tau_\lambda^{\sigma\mu}}\right)\\
        &= v_{\sigma\tau}\left(\frac{1}{2}g^{\mu\lambda}\epsilon^{\sigma\nu\alpha\beta}
        \tensor{R}{^\tau _{\lambda\alpha\beta}}
        + \frac{1}{2}g^{\nu\lambda}\epsilon^{\sigma\mu\alpha\beta}
        \tensor{R}{^\tau _{\lambda\alpha\beta}}\right)\\
        &= \frac{1}{2}v_{\sigma\tau}\tensor{R}{^\tau _{\lambda\alpha\beta}}
        \left[\left(\Bar{g}^{\mu\lambda}\epsilon^{\sigma\nu\alpha\beta}
        + \Bar{g}^{\nu\lambda}\epsilon^{\sigma\mu\alpha\beta} \right)
        + \epsilon\left(h^{\mu\lambda}\epsilon^{\sigma\nu\alpha\beta}
        + h^{\nu\lambda}\epsilon^{\sigma\mu\alpha\beta}\right)\right].
    \end{aligned}
\end{equation}
We can expand the Riemann tensor components as
\begin{equation}
    \tensor{R}{^\tau _{\lambda\alpha\beta}} = \tensor{\Bar{R}}{^\tau _{\lambda\alpha\beta}}
    + \epsilon\,\delta\tensor{R}{^\tau _{\lambda\alpha\beta}},
\end{equation}
and using this we get
\begin{equation}
    C^{\mu\nu}_{(2)} = \Bar{C}^{\mu\nu}_{(2)} +\frac{\epsilon}{2}
    \left[v_{\sigma\tau}\delta\tensor{R}{^\tau _{\lambda\alpha\beta}}
    \left(\Bar{g}^{\mu\lambda}\epsilon^{\sigma\nu\alpha\beta}
    + \Bar{g}^{\nu\lambda}\epsilon^{\sigma\mu\alpha\beta} \right)\right] + \mathcal{O}(\epsilon^2),
\end{equation}
which may be also written as
\begin{equation}
    \delta C^{(2)}_{\mu\nu} = \frac{1}{2}v_{\sigma\tau}
    \left(\Bar{g}_{\nu\gamma}\epsilon^{\sigma\gamma\alpha\beta}
    \delta\tensor{R}{^\tau _{\mu\alpha\beta}}
    + \Bar{g}_{\mu\rho}\epsilon^{\sigma\rho\alpha\beta}\delta\tensor{R}{^\tau _{\nu\alpha\beta}}\right).
\end{equation}
Hence the first-order perturbation term, when we denote the full Cotton tensor as
\begin{equation}
    C_{\mu\nu} = \Bar{C}_{\mu\nu} + \epsilon \delta C_{\mu\nu} + \mathcal{O}(\epsilon^2),
\end{equation}
may be expressed as
\begin{equation}
    \delta C_{\mu\nu} = \left[v_{\alpha}\left(\epsilon^{\alpha\beta\sigma\tau}
    \Bar{g}_{\mu\beta}\nabla_{\sigma}\delta R_{\nu\tau}
    + \epsilon^{\alpha\gamma\sigma\tau}\Bar{g}_{\nu\gamma}\nabla_{\sigma}\delta R_{\mu\tau}\right)
    + \frac{1}{2}v_{\sigma\tau}\left(\Bar{g}_{\nu\gamma}\epsilon^{\sigma\gamma\alpha\beta}
    \delta\tensor{R}{^\tau _{\mu\alpha\beta}} + \Bar{g}_{\mu\rho}\epsilon^{\sigma\rho\alpha\beta}
    \delta\tensor{R}{^\tau _{\nu\alpha\beta}}\right)\right].
\end{equation}

We can also look at the perturbed form of the electromagnetic stress-energy tensor, due to changes in the
metric structure. The stress-energy tensor for the electromagnetic field is
\begin{equation}
\begin{aligned}
    T_{\mu\nu} &=  F_{\mu\alpha}g^{\alpha\beta}F_{\beta\nu}
    - \frac{1}{4}g_{\mu\nu}F_{\sigma\alpha}g^{\alpha\beta}F_{\beta\rho}g^{\rho\sigma}\\
    &= F_{\mu\alpha}F^{\alpha}{}_{\nu} - \frac{1}{4}g_{\mu\nu}F^2,
\end{aligned}
\end{equation}
where $F^2 = F_{\sigma\alpha}F^{\sigma\alpha}$\\
Using the form of the perturbed metric, we can calculate the perturbations:
\begin{equation}
\begin{aligned}
    T_{\mu\nu} = &\, F_{\mu\alpha}\bar{g}^{\alpha\beta}F_{\nu\beta}
    + \epsilon F_{\mu\alpha}h^{\alpha\beta}F_{\nu\beta} - \frac{1}{4}\left(\bar{g}_{\mu\nu}
    + \epsilon h_{\mu\nu}\right)
    \left[F_{\sigma\alpha}\bar{g}^{\alpha\beta}F_{\beta\rho}\bar{g}^{\rho\sigma}\right.\\
    &+ \left. \epsilon F_{\sigma\alpha}h^{\alpha\beta}F_{\beta\rho}\bar{g}^{\rho\sigma}
    + \epsilon F_{\sigma\alpha}\bar{g}^{\alpha\beta}F_{\beta\rho}h^{\rho\sigma}\right]
\end{aligned}
\end{equation}
If we denote the stress-energy tensor in the usual way as
\begin{equation}
    T_{\mu\nu} = \bar{T}_{\mu\nu} + \epsilon \delta T_{\mu\nu} + \mathcal{O}(\epsilon^2),
\end{equation}
then we find the perturbation in first order to be
\begin{equation}
    \delta T_{\mu\nu} = F_{\mu\alpha}h^{\alpha\beta}F_{\nu\beta} - \frac{1}{4}h_{\mu\nu}\bar{F}^2
    - \frac{1}{4}\bar{g}_{\mu\nu}F_{\sigma\alpha}\left[h^{\alpha\beta}F_{\beta\rho}\bar{g}^{\rho\sigma}
    + \bar{g}^{\alpha\beta}F_{\beta\rho}h^{\rho\sigma}\right],
\end{equation}
which may also be written
\begin{equation}
    \delta T_{\mu\nu} = -F_{\mu\alpha}\bar{g}^{\alpha\delta}h_{\delta\gamma}\bar{g}^{\beta\gamma}F_{\nu\beta}
    - \frac{1}{4}h_{\mu\nu}\bar{F}^2 + \frac{1}{4}\bar{g}_{\mu\nu}F_{\sigma\alpha}
    \left[\bar{g}^{\alpha\delta}h_{\delta\gamma}\bar{g}^{\beta\gamma}F_{\beta\rho}\bar{g}^{\rho\sigma}
    + \bar{g}^{\alpha\beta}F_{\beta\rho}\bar{g}^{\rho\delta}h_{\delta\gamma}\bar{g}^{\sigma\gamma}\right].
\end{equation}

\subsection{Diagonal Perturbations}


The general form taken by the first-order perturbations to the Einstein gravitational field tensor,
if we assume that the metric perturbations are only diagonal, is
\begin{equation}
    \delta G_{\mu\nu} = \frac{1}{2}\left(\bar{g}^{\mu\lambda}
    \tensor{\bar{R}}{^\delta _{\lambda\nu\mu}}h_{\delta\mu} - \bar{R}h_{\mu\nu} 
    - \frac{1}{2}\bar{g}_{\mu\nu}\bar{g}^{\alpha\beta}\bar{g}^{\alpha\lambda}
    \tensor{\bar{R}}{^\delta _{\lambda\beta\alpha}}h_{\delta\alpha} 
    - \bar{g}_{\mu\nu}\bar{R}_{\alpha\beta}h^{\alpha\beta}\right).
\end{equation}
We may calculate how the perturbations affect specific components of the tensor, starting with $\delta G_{00}$:
\begin{equation}
    \begin{aligned}
        \delta G_{00} &= \frac{1}{2}\left(\bar{g}^{0\lambda}\tensor{\bar{R}}{^\delta _{\lambda 00}}
        h_{\delta 0} - \bar{R}h_{00} - \frac{1}{2}\bar{g}_{00}\bar{g}^{\alpha\beta}\bar{g}^{\alpha\lambda}
        \tensor{\bar{R}}{^\delta _{\lambda\beta\alpha}}h_{\delta\alpha}
        - \bar{g}_{00}\bar{R}_{\alpha\beta}h^{\alpha\beta}\right)\\
        &= \frac{1}{2}\left(\bar{g}^{00}\cancelto{0}{\tensor{\bar{R}}{^\delta _{0 00}}}h_{\delta 0}
        - \bar{R}h_{00} - \frac{1}{2}\bar{g}_{00}\bar{g}^{\alpha\beta}\bar{g}^{\alpha\lambda}
        \tensor{\bar{R}}{^\delta _{\lambda\beta\alpha}}h_{\delta\alpha}
        - \bar{g}_{00}\bar{R}_{\alpha\beta}h^{\alpha\beta}\right)\\
        &= \frac{1}{2}\left( - \bar{R}h_{00} - \frac{1}{2}\bar{g}_{00}\bar{g}^{\alpha\beta}
        \bar{g}^{\alpha\lambda}\tensor{\bar{R}}{^\delta _{\lambda\beta\alpha}}h_{\delta\alpha}
        - \bar{g}_{00}\bar{R}_{\alpha\beta}h^{\alpha\beta}\right).
        \label{eq-delta-G00}
    \end{aligned}
\end{equation}
For the index $\alpha = 0$, the term in parentheses receives the contribution
\begin{equation}
\begin{aligned}
    -\bar{R}h_{00} - \frac{1}{2}\bar{g}_{00}\bar{g}^{0\beta}\bar{g}^{0\lambda}\tensor{\bar{R}}
    {^\delta _{\lambda\beta 0}} h_{\delta 0} - \bar{g}_{00}\bar{R}_{0\beta}h^{0\beta}
     =& -\bar{R}h_{00} - \frac{1}{2}\bar{g}_{00}\bar{g}^{00}\bar{g}^{00}\cancelto{0}{\tensor{\bar{R}}
     {^\delta _{000}}} h_{\delta 0} - \bar{g}_{00}\bar{R}_{0\beta}h^{0\beta}\\
     =& -\bar{R}h_{00} + \bar{g}_{00}\bar{R}^{00}h_{00}.
\end{aligned}
\end{equation}
On the other hand, with a spacelike index $\alpha = i$, the contribution to the expression in parentheses
in (\ref{eq-delta-G00}) is
\begin{equation}
\begin{aligned}
    -\bar{R}h_{00} - \frac{1}{2}\bar{g}_{00}\bar{g}^{i\beta}\bar{g}^{i\lambda}\tensor{\bar{R}}
    {^\delta _{\lambda\beta i}} h_{\delta i} - \bar{g}_{00}\bar{R}_{i\beta}h^{i\beta}
     =& -\bar{R}h_{00} - \frac{1}{2}\bar{g}_{00}\bar{g}^{ii}\bar{g}^{jj}\cancelto{0}{\tensor{\bar{R}}
     {^\delta _{iii}}} h_{\delta i} - \bar{g}_{00}\bar{R}_{i\beta}h^{i\beta}\\
     =& -\bar{R}h_{00} + \bar{g}_{00}\bar{R}^{ii}h_{jj}.
\end{aligned}
\end{equation}
So the total first-order perturbation to the time-time component is
\begin{equation}
\begin{aligned}
    \delta G_{00} &= \frac{1}{2}\left[-\bar{R}h_{00} + \bar{g}_{00}\left(\bar{R}^{00}h_{00}
    + \bar{R}^{ii}h_{jj}\right)\right]\\
    &= \frac{1}{2}\left[-\left(\bar{g}_{00}\bar{R}^{00} + \bar{g}_{ii}\bar{R}^{jj}\right)h_{00}
    + \bar{g}_{00}\left(\bar{R}^{00}h_{00} + \bar{R}^{ii}h_{jj}\right)\right]\\
    &= \frac{1}{2}\left(\bar{g}_{00}\bar{R}^{ii}h_{jj} - \bar{g}_{ii}\bar{R}^{jj}h_{00}\right).
\end{aligned}
\end{equation}
When there are spatial indices in involved, the perturbations may be calculated the using the same method.
For a fixed index $i$ (not summed), the results are
\begin{equation}
    \begin{aligned}
        \delta G_{ii} &= \frac{1}{2}\left[\bar{g}_{ii}\left(\bar{R}^{00}h_{00} + \bar{R}^{jj}h_{kk}\right)
        - \left(\bar{g}_{00}\bar{R}^{00} + \bar{g}_{jj}\bar{R}^{jj}\right)h_{ii}\right]\\
        \delta G_{0i} &= \frac{1}{2}\left(\bar{g}^{00}\tensor{\bar{R}}{^\delta _{0i0}}h_{\delta 0}\right) = 0\\
        \delta G_{ij} &= \frac{1}{2}\left(\bar{g}^{ii}\tensor{\bar{R}}{^\delta _{iji}}h_{\delta i}\right) = 0,
    \end{aligned}
\end{equation}
with the last expression also involving another fixed spatial index $j\neq i$.

In a similar vein, we may find the perturbed form of the Cotton tensor. It is
\begin{equation}
    \begin{aligned}
        \delta C_{\mu\nu} = &\,\frac{1}{2} \Bigg[ \Big( v_{\alpha} \epsilon^{\alpha\gamma\sigma\tau}
        \bar{g}_{\nu\gamma} \bar{g}^{\mu\lambda} \nabla_{\sigma} \tensor{\bar{R}}{^\delta_{\lambda\mu\tau}}
        + \frac{1}{2}v_{\sigma\tau} \bar{g}_{\nu\gamma} \bar{g}^{\mu\tau}
        \epsilon^{\sigma\gamma\alpha\beta} \tensor{\bar{R}}{^\delta_{\mu\alpha\beta}} \Big) h_{\delta\mu} \\
       & + \Big( v_{\alpha} \epsilon^{\alpha\beta\sigma\tau} \bar{g}_{\mu\beta} \bar{g}^{\nu\lambda}
        \nabla_{\sigma} \tensor{\bar{R}}{^\delta_{\lambda\nu\tau}} + \frac{1}{2}v_{\sigma\tau}
        \bar{g}_{\mu\rho} \bar{g}^{\nu\tau} \epsilon^{\sigma\rho\alpha\beta}
        \tensor{\bar{R}}{^\delta_{\nu\alpha\beta}} \Big) h_{\delta\nu}\\
       & + v_{\alpha}\left(\epsilon^{\alpha\beta\sigma\tau}\bar{g}_{\mu\beta}\bar{g}^{\nu\lambda}
       \tensor{\bar{R}}{^\delta_{\lambda\nu\tau}}\nabla_{\sigma}h_{\delta\nu}
       + \epsilon^{\alpha\gamma\sigma\tau}\bar{g}_{\nu\gamma}\bar{g}^{\mu\lambda}
       \tensor{\bar{R}}{^\delta_{\lambda\mu\tau}}\nabla_{\sigma}h_{\delta\mu}\right)\Bigg].
    \end{aligned}
\end{equation}
We use this general expression to calculate the form of the time-time component, yielding
\begin{equation}
        \delta C_{00} =  \Bigg[ \Big( v_{\alpha} \epsilon^{\alpha\beta\sigma\tau} \bar{g}_{0\beta}
        \bar{g}^{0\lambda} \nabla_{\sigma} \tensor{\bar{R}}{^\delta_{\lambda 0\tau}}
        + \frac{1}{2}v_{\sigma\tau} \bar{g}_{0\rho} \bar{g}^{0\tau} \epsilon^{\sigma\rho\alpha\beta}
         \tensor{\bar{R}}{^\delta_{0\alpha\beta}} \Big) h_{\delta 0}
       + v_{\alpha}\left(\epsilon^{\alpha\beta\sigma\tau}\bar{g}_{0\beta}\bar{g}^{0\lambda}\tensor{\bar{R}}
       {^\delta_{\lambda 0\tau}}\nabla_{\sigma}h_{\delta 0}\right)\Bigg],
\end{equation}
which, since the background metric is diagonal, simplifies to
\begin{equation}
        \delta C_{00} =  \Bigg[ \Big( v_{\alpha} \epsilon^{\alpha 0\sigma\tau}\nabla_{\sigma}
        \tensor{\bar{R}}{^\delta_{0 0\tau}} + \frac{1}{2}v_{\sigma 0}\epsilon^{\sigma 0\alpha\beta}
         \tensor{\bar{R}}{^\delta_{0\alpha\beta}} \Big) h_{\delta 0} 
       + v_{\alpha}\left(\epsilon^{\alpha 0\sigma\tau}\tensor{\bar{R}}{^\delta_{0 0\tau}}
       \nabla_{\sigma}h_{\delta 0}\right)\Bigg]=0.
\end{equation}
The further assumption that the perturbation is itself diagonal implies that the index $\delta$ has to be
zero for this term to be potentially nonvanishing. However, even in this case, we find that the
final expression involves the Riemann tensor with its first two indices identical,
\begin{equation}
    \delta C_{00} = \frac{1}{2}\left(v_{\sigma 0}\epsilon^{\sigma 0\alpha\beta}
    \tensor{\bar{R}}{^0 _{0\alpha\beta}}\right)h_{00}=0.
\end{equation}
Again, by a similar method we can calculate the perturbation terms in the other components also. For
fixed $i$,
\begin{equation}
    \begin{aligned}
        \delta C_{ii} &= \frac{1}{2}\left(v_{\sigma i}\epsilon^{\sigma i \alpha\beta}
        \tensor{\bar{R}}{^i _{i\alpha\beta}}\right)h_{ii}=0\\
        \delta C_{0i} &= \frac{1}{2}\left(v_{\sigma 0}\bar{g}_{ii}\bar{g}^{00}\epsilon^{\sigma i\alpha\beta}
        \tensor{\bar{R}}{^0 _{0\alpha\beta}}h_{00} + v_{\sigma i}\bar{g}_{00}\bar{g}^{ii}
        \epsilon^{\sigma 0\alpha\beta}\tensor{\bar{R}}{^i _{i\alpha\beta}}h_{ii}\right)=0\\
        \delta C_{ij} &= \frac{1}{2}\left(v_{\sigma i}\bar{g}_{jj}\bar{g}^{ii}
        \epsilon^{\sigma j\alpha\beta}\tensor{\bar{R}}{^i _{i\alpha\beta}}h_{ii} + v_{\sigma j}
        \bar{g}_{ii}\bar{g}^{jj}\epsilon^{\sigma i\alpha\beta}\tensor{\bar{R}}{^j _{j\alpha\beta}}h_{jj}\right)
        =0,
    \end{aligned}
\end{equation}
all vanishing in this regime, as expected based on previous work on diagonal perturbations in spherically
symmetric spacetimes.

For the Maxwell stress-energy tensor, the perturbation has the general form,
\begin{equation}
    \delta T_{\mu\nu} = -F_{\mu\alpha}\bar{g}^{\alpha\delta}h_{\delta\gamma}\bar{g}^{\beta\gamma}F_{\nu\beta}
    - \frac{1}{4}h_{\mu\nu}\bar{F}^2 + \frac{1}{4}\bar{g}_{\mu\nu}F_{\sigma\alpha}
    \left(\bar{g}^{\alpha\delta}h_{\delta\gamma}\bar{g}^{\beta\gamma}F_{\beta\rho}\bar{g}^{\rho\sigma}
    + \bar{g}^{\alpha\beta}F_{\beta\rho}\bar{g}^{\rho\delta}h_{\delta\gamma}\bar{g}^{\sigma\gamma}\right).
\end{equation}
The standard four-vector potential for a charged BH is
\begin{equation}
    \bar{A}^{\mu} = \left(\frac{Q}{r},0,0,0\right),
\end{equation}
which gives only one component of the field strength tensor ($F_{\mu\nu}$) which is
the radial electrostatic field $E_{r}=F_{tr}=F_{01} = -F_{10}$.
Using this fact we can calculate the perturbation in the time-time component, the energy density,
\begin{equation}
    \begin{aligned}
        \delta T_{00} &= -F_{0\alpha}\bar{g}^{\alpha\delta}h_{\delta\gamma}\bar{g}^{\beta\gamma}F_{0\beta}
        - \frac{1}{4}h_{00}\bar{F}^2 + \frac{1}{4}\bar{g}_{00}F_{\sigma\alpha}
        \left(\bar{g}^{\alpha\delta}h_{\delta\gamma}\bar{g}^{\beta\gamma}F_{\beta\rho}\bar{g}^{\rho\sigma}
        + \bar{g}^{\alpha\beta}F_{\beta\rho}\bar{g}^{\rho\delta}h_{\delta\gamma}\bar{g}^{\sigma\gamma}\right)\\
        &= -F_{01}\bar{g}^{1\delta}h_{\delta\gamma}\bar{g}^{\beta 1}F_{01} - \frac{1}{4}h_{00}\bar{F}^2
        + \frac{1}{4}\bar{g}_{00}F_{\sigma\alpha}\left(\bar{g}^{\alpha\delta}h_{\delta\gamma}
        \bar{g}^{\beta\gamma}F_{\beta\rho}\bar{g}^{\rho\sigma} + \bar{g}^{\alpha\beta}F_{\beta\rho}
        \bar{g}^{\rho\delta}h_{\delta\gamma}\bar{g}^{\sigma\gamma}\right]\\
        &= -F_{01}\bar{g}^{11}h_{11}\bar{g}^{11}F_{01} - \frac{1}{4}h_{00}\bar{F}^2
        + \frac{1}{4}\bar{g}_{00}F_{\sigma\alpha}\left(\bar{g}^{\alpha\delta}h_{\delta\gamma}
        \bar{g}^{\beta\gamma}F_{\beta\rho}\bar{g}^{\rho\sigma} + \bar{g}^{\alpha\beta}F_{\beta\rho}
        \bar{g}^{\rho\delta}h_{\delta\gamma}\bar{g}^{\sigma\gamma}\right).
        \label{eq-delta-T00}
    \end{aligned}
\end{equation}
The last term in the above expression can be calculated in two steps,
again separating out the timelike index $\sigma = 0$ from spacelike $\sigma = 1$. For $\sigma = 0$ we get
\begin{equation}
\begin{aligned}
    \frac{1}{4}\bar{g}_{00}F_{0\alpha}\left(\bar{g}^{\alpha\delta}h_{\delta\gamma}
    \bar{g}^{\beta\gamma}F_{\beta\rho}\bar{g}^{\rho 0} + \bar{g}^{\alpha\beta}F_{\beta\rho}
    \bar{g}^{\rho\delta}h_{\delta\gamma}\bar{g}^{0\gamma}\right)
    &=\frac{1}{4}\bar{g}_{00}F_{01}\left(\bar{g}^{1\delta}h_{\delta\gamma}\bar{g}^{\beta\gamma}
    F_{\beta\rho}\bar{g}^{\rho 0} + \bar{g}^{1\beta}F_{\beta\rho}\bar{g}^{\rho\delta}
    h_{\delta\gamma}\bar{g}^{0\gamma}\right)\\
    &=\frac{1}{4}\bar{g}_{00}F_{01}\left(\bar{g}^{11}h_{11}\bar{g}^{11}F_{10}\bar{g}^{ 00}
    + \bar{g}^{11}F_{10}\bar{g}^{00}h_{00}\bar{g}^{00}\right).
\end{aligned}
\end{equation}
When $\sigma = 1$, the analogous calculation yields
\begin{equation}
    \begin{aligned}
         \frac{1}{4}\bar{g}_{00}F_{1\alpha}\left(\bar{g}^{\alpha\delta}h_{\delta\gamma}
         \bar{g}^{\beta\gamma}F_{\beta\rho}\bar{g}^{\rho 1} + \bar{g}^{\alpha\beta}F_{\beta\rho}
         \bar{g}^{\rho\delta}h_{\delta\gamma}\bar{g}^{1\gamma}\right)
         = \frac{1}{4}\bar{g}_{00}F_{10}\left(\bar{g}^{00}h_{00}\bar{g}^{00}F_{01}\bar{g}^{11}
         + \bar{g}^{00}F_{01}\bar{g}^{11}h_{11}\bar{g}^{11}\right).
    \end{aligned}
\end{equation}
This sum of these terms, when simplified, may be written as
\begin{equation}
\begin{aligned}
    \frac{1}{4}(\bar{g}^{11}\bar{g}^{11}F_{01}F_{10}h_{11}
    + \bar{g}^{11}\bar{g}^{00}F_{01}F_{10}h_{00} \, + & \, \bar{g}^{00}\bar{g}^{11}F_{10}F_{01}h_{00}
    + \bar{g}^{11}\bar{g}^{11}F_{10}F_{01}h_{11}) \\
    &= \frac{1}{2}\left(\bar{g}^{11}\bar{g}^{11}F_{01}F_{10}h_{11}
    + \bar{g}^{00}\bar{g}^{11}F_{10}F_{01}h_{00}\right) \\
    &= -\frac{1}{2}\left(\bar{g}^{11}\bar{g}^{11}h_{11}
    - \bar{g}^{00}\bar{g}^{11}h_{00}\right)(F_{01})^{2}
\end{aligned}
\end{equation}
Therefore, the complete form of the perturbation in the time-component becomes
\begin{equation}
    \delta T_{00} = -\frac{3}{2}F_{01}\bar{g}^{11}F_{01}\bar{g}^{11}h_{11} - \frac{1}{4}h_{00}\bar{F}^2
    - \frac{1}{2}\bar{g}^{11}F_{01}\bar{g}^{00}F_{01}h_{00}.
\end{equation}
Using the fact that $\bar{F}^2$, when there is only the radial electrostatic field, may be written as
\begin{equation}
    \begin{aligned}
        \bar{F}^2 &= F_{\sigma\alpha}\bar{g}^{\alpha\beta}F_{\beta\rho}\bar{g}^{\rho\sigma}\\
        &= F_{01}\bar{g}^{11}F_{10}\bar{g}^{00} + F_{10}\bar{g}^{00}F_{01}\bar{g}^{11}\\
        &= -2F_{01}\bar{g}^{00}F_{01}\bar{g}^{11},
    \end{aligned}
\end{equation}
we get the following form for the perturbation,
\begin{equation}
    \delta T_{00} = -\frac{3}{2}(\bar{g}^{11})^{2}(F_{01})^{2}h_{11}.
\end{equation}
The other terms in the perturbed stress-energy tensor [in $(r,theta,\phi)$ spherical coordinates] are similar,
\begin{equation}
    \begin{aligned}
        \delta T_{11} &= -\frac{3}{2}(\bar{g}^{00})^{2}(F_{01})^{2}h_{00} \\
        \delta T_{22} &= \frac{1}{2}\left[\bar{g}^{00}\bar{g}^{11}(F_{01})^{2}h_{22}
        + \bar{g}_{22}(\bar{g}^{11})^{2}\bar{g}^{00}(F_{01})^{2}h_{11}
        + \bar{g}_{22}(\bar{g}^{00})^{2}\bar{g}^{11}(F_{01})^{2}h_{00}\right]\\
        \delta T_{33} &= \frac{1}{2}\left[\bar{g}^{00}\bar{g}^{11}(F_{01})^{2}h_{33}
        + \bar{g}_{33}(\bar{g}^{11})^{2}\bar{g}^{00}(F_{01})^{2}h_{11}
        + \bar{g}_{33}(\bar{g}^{00})^{2}(F_{01})^{2}\bar{g}^{11}h_{00}\right].
    \end{aligned}
\end{equation}
These are all naturally proportional to the square of the electric field strength, and all the
other (off-diagonal) perturbation terms have been calculated to be nonexistent.

\subsection{Azimuthal Perturbations}

We should also consider the principal case treated in the body of this paper---that of a perturbation
to the metric that only has a $t$- and $\phi$-independent $h_{t\phi}=H(r,\theta)$. The calculations starting
from (\ref{eq-delta-R}) follow along similar lines to what we saw in the case of an arbitrary diagonal
metric perturbation, although obviously with different surviving nonzero terms.
The only component of the Ricci tensor which picks up an extra term at first order in $\epsilon$ is
the $R_{t\phi}$ term,
\begin{equation}
    \delta R_{t\phi} = \frac{f}{2}\partial^2_r H + \frac{1}{2r^2}\partial^2_{\theta}H
    - \frac{\cot{\theta}}{r^2}\partial_{\theta}H + \frac{f'}{r}H.
\end{equation}
All the other new Ricci tensor components follow the same pattern as the metric perturbation $h_{\mu\nu}$
and vanish. The perturbations to the Cotton tensor calculated for the $t$-$\phi$ coordinates are also zero.
The modification to the Einstein tensor is
\begin{equation}
\begin{aligned}
    G_{t\phi}=\delta G_{t\phi} &= \delta R_{t\phi} - \frac{1}{2}g_{t\phi}\bar{R} \\
    &= \frac{f}{2}\partial^2_r H + \frac{1}{2r^2}\partial^2_{\theta}H - \frac{\cot{\theta}}{r^2}\partial_{\theta}H
     + \left(\frac{1}{r^2} - \frac{f}{r^2} - \frac{f'}{r} - \frac{f''}{2}\right)H\\
    &= \frac{f}{2}\partial^2_r H + \frac{1}{2r^2}\partial^2_{\theta}H - \frac{\cot{\theta}}{r^2}\partial_{\theta}H
     + \left(\frac{2M}{r^3} - \frac{2Q^2}{r^4}\right)H
\end{aligned}
\end{equation}
We can further show that the only term which has a perturbation up to first order in the
electromagnetic stress-energy is likewise the $T_{t\phi}$ term, which is
\begin{equation}
    T_{t\phi} = -\frac{1}{4\pi}\frac{Q^2}{r^4}H.
\end{equation}
Hence the Einstein equation for the $t$-$\phi$ term takes the form,
\begin{equation}
    \frac{f}{2}\partial^2_r H + \frac{1}{2r^2}\partial^2_{\theta}H
    - \frac{\cot{\theta}}{r^2}\partial_{\theta}H + \left(\frac{2M}{r^3} - \frac{2Q^2}{r^4}\right)H
     = -\frac{2Q^2}{r^4}H.
\end{equation}
With a bit of further simplification using the explicit form of $f(r) = 1 - 2M/r + Q^2/r^2$,
we see that the this equation reduces to
\begin{equation}
     \frac{r^2f}{2}\partial^2_r H + \frac{2M}{r}H + \frac{1}{2}\partial^2_{\theta}H
     - \cot{\theta}\,\partial_{\theta}H = 0
\end{equation}

\subsection{Kretschmann Scalar}

The Kretschmann scalar is defined to be
\begin{equation}
    K = \tensor{R}{^{\mu\nu\rho\sigma}}\tensor{R}{_{\mu\nu\rho\sigma}}.
\end{equation}
We shall evaluate this in the perturbed spacetime with the azimuthal $h_{t\phi}$.
In the case of a perturbed Riemann tensor of the form
\begin{equation}
    \tensor{R}{^\mu_{\nu\rho\sigma}} = \tensor{\bar{R}}{^\mu_{\nu\rho\sigma}}
    + \epsilon\delta\tensor{R}{^\mu_{\nu\rho\sigma}},
\end{equation}
upon substitution we get the form of the scalar as follows:
\begin{equation}
\begin{aligned}
    K &= (\tensor{\bar{R}}{^{\mu\nu\rho\sigma}} + \epsilon\delta\tensor{R}{^{\mu\nu\rho\sigma}})
    (\tensor{\bar{R}}{_{\mu\nu\rho\sigma}} + \epsilon\delta\tensor{R}{_{\mu\nu\rho\sigma}})\\
    &= \bar{K} + 2\epsilon\tensor{\bar{R}}{^{\mu\nu\rho\sigma}}\delta\tensor{R}{_{\mu\nu\rho\sigma}}
    + \mathcal{O}(\epsilon^2)
\end{aligned}
\end{equation}
So the extra term becomes a product of the background metric terms and the perturbation associated with the
introduction of the off-diagonal term in the metric. For the background metric in the original
orthonormal basis we have calculated that no perturbations are introduced into the originally nonzero Riemann
tensor components; rather, the perturbation only makes some previously vanishing components nonzero.
So the full contraction of the background curvature tensor and the curvature perturbations is
vanishing, and the Kretschmann scalar stays the same as it was with only the background metric,
\begin{equation}
    K = \bar{K} = \frac{48M^2}{r^6} - \frac{96MQ^2}{r^7} + \frac{56Q^4}{r^8}.
\end{equation}

\section{Appendix: WKB Analysis}
\label{app:wkb}

In this appendix, we examine more closely the radial profile of the perturbations
to the $h_{t\phi}$ term in the metric, focusing on the regime with a large angular parameter, $l \geq 16$. 
The radial perturbation equation takes the form,
\begin{equation}
\frac{d^2 R}{dr^2} + \left[\frac{4M}{f r^3} - \frac{2l(l+1)}{f r^2}\right]R = 0,
\label{eq:main}
\end{equation}
where $f(r) = 1 - 2M/r + Q^2/r^2$. For large $l$ values, this equation becomes highly oscillatory,
making the WKB approximation particularly well suited for our analysis.

\subsection{WKB Setup}

Rewriting (\ref{eq:main}) in Schr\"{o}dinger form,
\begin{equation}
\frac{d^2 R}{dr^2} + k^2(r)R = 0,
\end{equation}
the local effective wave number is
\begin{equation}
\label{eq-eff-wave-num}
k^2(r) = \frac{4M}{f r^3} - \frac{2l(l+1)}{f r^2}.
\end{equation}
Between the horizons ($r_- < r < r_+$), where $f(r) < 0$, the potential becomes
\begin{equation}
k^2(r) \approx \frac{2l^2}{|f| r^2} \left(1 - \frac{2M}{l^2 r}\right).
\end{equation}

The WKB approximation requires that the wavelength $\lambda(r) = 2\pi/k(r)$ vary slowly compared to
the local scale---that is, that $|dk/dr| \ll k^2$. Near the horizons $r \to r_{\pm}$, where $f(r) \to 0$, the
effective wave number $k(r)$ from \eqref{eq-eff-wave-num} actually diverges as $|f(r)|^{-1/2}$,
meaning $k(r) \to \infty$. Consequently, the wavelength vanishes and the adiabatic condition
$|dk/dr| \ll k^2$ is actually satisfied arbitrarily close to the horizons, as the rapid oscillations
validate the WKB ansatz. For resonance modes, a perturbation's amplitude through most of the
$r_{-}<r<r_{+}$ region is very large compared with its values at the horizon boundaries.
This justifies the use of the version
(\ref{eq:bohr_quantization}) of the Bohr-Sommerfeld quantization condition
corresponding to hard walls at the boundaries, and
which connects the two horizon boundaries through the oscillatory region.

The second-order WKB solution takes the form
\begin{equation}
R(r) \approx \frac{A_{\textrm{WKB}}}{\sqrt{k(r)}}\cos\left(\int dr\, k(r) + \varphi\right),
\label{eq:wkb_sol}
\end{equation}
where $\varphi$ is a phase factor. [Note that, although the overall normalization of the solution is
essentially arbitrary when using this method, there are suggestions that there may be suppression
of the amplitude at larger values of $l$, because of the factor of $\sqrt{k(x)}$ in the denominator
of the prefactor of (\ref{eq:wkb_sol}).]
If the perturbation goes to zero on either side outside this region,
there is an approximate quantization condition following from
\begin{equation}
\label{eq:bohr_quantization}
\int_{r_-}^{r_+}dr\, k(r) = \left(n +1\right)\pi,
\end{equation}
for $n \in \mathbb{Z}^{+}$.

For $l \gg 1$, the solution simplifies to
\begin{equation}
R(r) \propto \sqrt{r}|f|^{1/4}\cos\left(l\sqrt{2}\int_{r_{-}}^{r_{+}}\frac{dr}{r\sqrt{|f|}} + \varphi\right),
\end{equation}
with the integral evaluating to
\begin{equation}
I=\int_{r_{-}}^{r_{+}} \frac{dr}{r\sqrt{|f|}} = \int \frac{r\,dr}{\sqrt{(r_+-r)(r-r_-)}},
\end{equation}
which is a solvable elliptic integral, using a change of variables,
\begin{equation}
    r = r_- + (r_+ - r_-)\sin^2{\psi}.
\end{equation}
In terms of the new integration variable $\psi$,
\begin{equation}
    \begin{aligned}
        dr &= 2(r_+ - r_-)\sin{\psi}\cos{\psi}\,d\psi\\
        r_+ - r &= (r_+ - r_-)\cos^2{\psi}\\
        r - r_- &= (r_+ - r_-)\sin^2{\psi}.
    \end{aligned}
\end{equation}
Using this, the integral becomes:
\begin{equation}
    I = \int_{0}^{\pi/2}\frac{[r_- + (r_+ - r_-)\sin^2{\psi}] \,
    2(r_+ - r_-)\sin{\psi}\cos{\psi}\,d\psi}{(r_+ - r_-)\sin{\psi}\cos{\psi}},
\end{equation}
which simplifies to
\begin{equation}
I = 2\int_0^{\pi/2}d\psi \left[\frac{M}{\sqrt{M^2 - Q^2}} - 1 + 2\sin^2\psi\right].
\label{eq:simplified}
\end{equation}

\subsection{Exact Evaluation}

Since the integrals now have the elementary values,
\begin{equation}
\begin{aligned}
\int_0^{\pi/2} d\theta &= \frac{\pi}{2} \\
\int_0^{\pi/2} \sin^2\theta\,d\theta &= \frac{\pi}{4},
\end{aligned}
\end{equation}
the whole expression for $I$ becomes
\begin{equation}
I = 2\left(\frac{M\pi}{2\sqrt{M^2 - Q^2}} - \frac{\pi}{2} + \frac{\pi}{2}\right) =
\frac{\pi M}{\sqrt{M^2 - Q^2}},
\label{eq:finalI}
\end{equation}
making the phase condition
\begin{equation}
l\sqrt{2}\frac{\pi M}{\sqrt{M^2 - Q^2}} = \left(n + 1\right)\pi.
\end{equation}
This, in turn, simplifies to the analytic quantization rule for the resonant modes,
\begin{equation}
\frac{lM\sqrt{2}}{\sqrt{M^2 - Q^2}} = n + 1.
\label{eq:quantization}
\end{equation}
The possible resonant values of $n$ evidently depend on $l$, $M$, and $\sqrt{M^{2}-Q^{2}}$. Increasing $Q$
decreases the number of allowed nodes (all other parameters being kept the same). In contrast,
increasing $l$ increases the number of nodes, something we clearly saw in the numerical data.
The exact WKB solution between horizons is
\begin{equation}
R(r) = A_{\textrm{WKB}}\left[\frac{r}{\sqrt{|f(r)|}}\right]^{1/2}\!
\cos\left[\frac{lM\sqrt{2}}{\sqrt{M^2 - Q^2}}\sin^{-1}\left(\sqrt{\frac{r - r_-}{r_+ - r_-}}\right)
+ \varphi\right].
\label{eq:solution}
\end{equation}

\subsection{Limiting Cases}

In the Schwarzschild limit, the WKB method produces
a reasonably satisfactory approximate analytic solution. For $Q \rightarrow 0$, the integral
$I\rightarrow\pi$, leaving the quantization condition as
\begin{equation}
\frac{l}{\sqrt{2}} \approx n + 1,
\end{equation}
which is interesting, in that only for certain matched values of $l$ and $n$ can it be satisfied. Unlike
in a quantum mechanics problem, there is not a state for every non-negative integer $n$.

In the extremal limit, $I\rightarrow\infty$, and the solution breaks down---although this may not be
so surprising, since $r_{-}$ and $r_{+}$ approach the same value in that limit, leaving a very limited
spatial region over which to calculate the WKB phase.

\end{document}